\documentclass[twocolumn]{aastex62}
\usepackage{mathtools}
\usepackage[maxfloats=40]{morefloats} 
\usepackage{color}
\usepackage{amsmath}
\usepackage[figure,figure*]{hypcap}

\DeclarePairedDelimiter\abs{\lvert}{\rvert}%
\DeclarePairedDelimiter\norm{\lVert}{\rVert}%

\makeatletter
\let\oldabs\abs
\def\abs{\@ifstar{\oldabs}{\oldabs*}}
\let\oldnorm\norm
\def\norm{\@ifstar{\oldnorm}{\oldnorm*}}
\makeatother

\newcommand{\Rom}[1]{\uppercase\expandafter{\romannumeral #1\relax}}
\def \jyb {Jy~beam$^{-1}$}
\def \kms {km~s$^{-1}$}
\def\C18O{\textrm{C$^{18}$O}}

\begin{document}

\title{Formation of the Hub-Filament System G33.92+0.11: Local Interplay between Gravity, Velocity, and Magnetic Field}

\accepted{2020/11/3}
\submitjournal{\apj}

\author[0000-0002-6668-974X]{Jia-Wei Wang}
\affiliation{Academia Sinica Institute of Astronomy and Astrophysics, No.1, Sec. 4., Roosevelt Road, Taipei 10617, Taiwan}
\affiliation{Institute of Astronomy and Department of Physics, National Tsing Hua University, Hsinchu 30013, Taiwan}
\email{jwwang@asiaa.sinica.edu.tw}

\author[0000-0003-2777-5861]{Patrick M. Koch}
\affiliation{Academia Sinica Institute of Astronomy and Astrophysics, No.1, Sec. 4., Roosevelt Road, Taipei 10617, Taiwan}

\author[0000-0003-1480-4643]{Roberto Galv{\'a}n-Madrid}
\affiliation{Instituto de Radioastronom{\'i}a y Astrof{\'i}sica, Universidad Nacional Aut{\'o}noma de M{\'e}xico, Apdo. Postal 3-72 (Xangari), 58089 Morelia, Michoac{\'a}n, Mexico}

\author[0000-0001-5522-486X]{Shih-Ping Lai}
\affiliation{Institute of Astronomy and Department of Physics, National Tsing Hua University, Hsinchu 30013, Taiwan}

\author[0000-0003-2300-2626]{Hauyu Baobab Liu}
\affiliation{Academia Sinica Institute of Astronomy and Astrophysics, No.1, Sec. 4., Roosevelt Road, Taipei 10617, Taiwan}

\author[0000-0002-6868-4483]{Sheng-Jun Lin}
\affiliation{Institute of Astronomy and Department of Physics, National Tsing Hua University, Hsinchu 30013, Taiwan}

\author[0000-0002-8557-3582]{Kate Pattle}
\affiliation{Centre for Astronomy, School of Physics, National University of Ireland Galway, University Road, Galway, Ireland}
\affiliation{Institute of Astronomy and Department of Physics, National Tsing Hua University, Hsinchu 30013, Taiwan}


\begin{abstract}
The formation of filaments in molecular clouds is an important process in star formation.
Hub-filament systems (HFSs) are a transition stage connecting parsec-scale filaments and proto-clusters. Understanding the origin of HFSs is crucial to reveal how star formation proceeds from clouds to cores. Here, we report JCMT POL-2 850 $\mu$m polarization and IRAM 30-m C$^{18}$O (2-1) line observations toward the massive HFS G33.92+0.11. 
The 850 $\mu$m continuum map reveals four major filaments converging to the center of G33.92+0.11 with numerous short filaments connecting to the major filaments at local intensity peaks. We estimate the local orientations of filaments, magnetic field, gravity, and velocity gradients from observations, and we examine their correlations based on their local properties. 
In the high-density areas, our analysis shows that the filaments tend to align with the magnetic field and local gravity. In the low-density areas, we find that the local velocity gradients tend to be perpendicular to both the magnetic field and local gravity, although the filaments still tend to align with local gravity. A global virial analysis suggests that the gravitational energy overall dominates the magnetic and kinematic energy. Combining local and global aspects, we conclude that the formation of G33.92+0.11 is predominantly driven by gravity, dragging
and aligning the major filaments and magnetic
field on the way to the inner dense center.
Traced by local velocity gradients in the outer diffuse areas, ambient gas might be accreted onto the major filaments directly or via the short filaments.


\end{abstract}


\section{Introduction}\label{sec:intro}
$Herschel$ observations of the Galactic interstellar medium revealed that filamentary structures are ubiquitous in molecular clouds, and that they are a key intermediate stage towards the formation of stars \citep[e.g.,][]{an10,ar11,an14}. Embedded in these filaments, many stars appear to form within clustered environments \citep[e.g.,][]{la03,gu09,ki13}. \citet{my09} reviewed previous observations and suggested that protoclusters are commonly associated with hub-filament systems (HFSs), where they are formed in a dense hub with numerous radial filaments extending from the central hub. Understanding how HFSs form is a topic of considerable interest, since HFSs are the possible transition stage connecting the evolution of filamentary clouds and the formation of protoclusters \citep[e.g.,][]{sc12}.

The origin of HFSs is still under debate. Observationally, HFS are seen across a variety of physical scales and environments, from the IRDC G335.43–0.24 with a physical size of $\sim5$ pc \citep{my09} to the $\sim0.3$-pc-sized HFS embedded within the Orion integral filament \citep{ha18}. As these HFSs are identified only based on their morphologies, the question remains whether all these systems form via similar mechanisms. Theoretical works suggest that HFS-like morphologies can be formed by a diversity of different mechanisms, such as layer fragmentation threaded by magnetic fields \citep[][]{my09,va14}, magnetic field channeled gravitational collapse \citep[e.g.,][]{na08}, shock/turbulence compression \citep[e.g.,][]{pa01a,fe16}, multi-scale gravitational collapse \citep[e.g.,][]{go14,go18}, and filament collisions \citep[e.g.,][]{na14,do14,fr15}. To further constrain these theories, it is crucial to observe key physical parameters related to these proposed physical mechanisms, such as velocity structures and magnetic fields.

Recent molecular line observations have revealed the kinematic structures of HFSs in nearby clouds. The line of sight (LOS) velocity gradient, traced by molecular lines, is commonly interpreted as a proxy for the plane-of-sky (POS) gas motion. Significant velocity gradients along filaments are seen in several HFSs \citep[e.g.,][]{fr13,liu12,ha18,ch20b}, suggesting that these filaments are likely dynamical gas flows and not static structures. The comparison between the density structures and the direction of the velocity gradient further suggests that these gas flows are possibly driven by gravity \citep[e.g.,][]{fr13,ha18}. Local velocity gradients within HFSs have been measured in a few systems \citep[e.g.,][]{pe14,wi18,ch20b}. These observations reveal a variation of velocity structures possibly influenced by shock compression \citep{wi18} or the balance between gas pressure and gravity \citep{ch20b}. Nevertheless, as molecular line observations only trace the LOS velocity, its gradient might also have a different origin, such as cloud collision, rather than the POS gas flows \citep{li19}. Hence, the alone analysis of LOS velocity gradients needs caution, and a joint analysis with additional measured physical 
parameters can lead to further insight \citep{ta19}.


Magnetic fields can be a key factor in the formation of molecular clouds and filaments, but their exact role is still uncertain \citep{li14}. Strong magnetic fields might constrain the gravitational collapse/fragmentation \citep[e.g.,][]{na08,my09,va14}, channel turbulent flows \citep[e.g.,][]{st08,li08}, and guide accreting gas \citep[e.g.,][]{an14,sh19}. At pc-scale, starlight polarization observations often reveal organized magnetic fields perpendicular to the longer axis of HFSs \citep[e.g.,][]{su11,sa16,wa20}, which has been interpreted as evidence for strong magnetic field models. However, other simulations suggest that the observed configurations can also be reproduced by mechanisms, such as multi-scale gravitational collapse \citep[e.g.,][]{go18} and shock compression \citep[e.g.,][]{ch20}. The recent JCMT BISTRO survey \citep{wa17} is probing the 1--0.01 pc scales magnetic fields toward numerous star-forming regions, including several known HFSs, such as IC5146 \citep{wa19}, NGC1333 \citep{do20}, NGC6334 \citep{ar20}. The BISTRO survey has revealed locally organized magnetic fields, displaying a variety of magnetic field morphologies around HFSs along filaments, possibly caused by a variable energy balance between gravity and magnetic field. These results suggest that the role of the magnetic field in HFSs is possibly scale-dependent and varies with the environments \citep[e.g.,][]{wa19,ar20}. However, polarization observations toward HFSs are still rare, and thus, more observational samples are still required to conclude on the role of magnetic fields in the formation of HFSs.

G33.92+0.11 is a remarkable ultracompact H\Rom{2} region that lies $7.1\substack{+1.2 \\ -0.3}$ kpc away in the Galactic plane \citep{fi03}. Since its derived virial mass (G33.92+0.11 A: 520 M$_{\sun}$; G33.92+0.11 B: 270 M$_{\sun}$) is much smaller than its enclosed molecular gas mass (G33.92+0.11 A: 5100 M$_{\sun}$; G33.92+0.11 B: 3200 M$_{\sun}$), it is likely geometrically thin and face-on \citep{wa99,liu12}. Thus, it is a good example to reveal the detailed morphology of an HFS. \citet{liu12} observed the 1.3 mm continuum emission within the inner 2 pc area of G33.92+0.11 and found that the system is composed of pc-long filaments converging to a massive ($3.0\substack{+2.8 \\ -1.4}\times10^3$ M$_{\sun}$) hub within the inner 0.6 pc, making this a typical HFS morphology. \citet{liu15} performed ALMA observations toward this central 0.6 pc area using several molecular lines. They found that this center contains several spiral-like filaments converging to two central massive ($100-300$ M$_{\sun}$) molecular cores. The 1000-au resolution 225 GHz continuum ALMA observations further identified 28 Class 0/I YSO candidates around these two massive cores, indicating that the center of this HFS is associated with cluster formation \citep{liu19}. These results reveal a picture where a massive protocluster can form within a central hub fed by gravitationally driven flows from 1-pc to 1000-au scales.

In this paper, we report the polarization and molecular line observations toward the G33.92+0.11 system, using the JCMT POL-2 polarimeter and the IRAM 30-m telescope. These observations, with a resolution of $\sim$0.5 pc, allow us to probe the magnetic field and kinematic structures of the G33.92+0.11 HFS at pc-scale. 
We aim at investigating how gravitational and magnetic fields interact with the density and kinematic structures in order to understand the formation of the G33.92+0.11 HFS. In \autoref{sec:obs}, we present the observations and data reduction. \autoref{sec:results} reports the observed magnetic field map and the kinematic structures of this system. \autoref{sec:ana} presents how we estimate the filamentary density structures, the local velocity gradient, and the local gravitational force from the observed data. 
A statistical analysis is performed to 
identify possible correlations between these physical parameters.
We discuss the dynamics of the system and its possible origin in \autoref{sec:discussion}. Our conclusions are summarized in \autoref{sec:con}.

\section{Observations}\label{sec:obs}
\subsection{JCMT POL-2 Observations}\label{sec:JCMT}
We carried out polarization continuum observations toward G33.92+0.11 with the reference position (R.A., Dec.)=(18$^{h}$52$^{m}$50.4$^{s}$, 00\degr55\arcmin28\farcs9) with SCUBA2 POL-2 mounted on the JCMT (project code M19AP034; PI: Jia-Wei Wang). Our target was observed with 24 sets of 40-minute integration between 2019 May 13 and 2019 June 2 under a $\tau_{225 GHz}$ opacity ranging from 0.03 to 0.06. 
The POL-2 DAISY scan mode \citep{fr16} was adopted, producing a fully sampled circular region with a 
diameter of 11\arcmin\ and a resolution of 14\arcsec,  
leading to a map with lowest and nearly uniform noise within the central 3 arcmin diameter region, and increasing noise towards the edge of the map. Both the 450 and 850 $\mu$m continuum polarization were observed simultaneously. 
This paper focuses on the 850 $\mu$m data.

The POL-2 polarization data were reduced using $pol2map$\footnote{http://starlink.eao.hawaii.edu/docs/sc22.pdf} in the \textsc{smurf} package\footnote{version 2019 Nov 2} \citep{be05,ch13}. The reduction procedure followed the $pol2map$ script. The $skyloop$ mode was invoked in order to reduce the uncertainty generated during the map-making process, which grows with intensities. The MAPVARS mode was activated, with which the uncertainty of the co-added images was calculated from the standard deviation among the individual observations, to account for the uncertainty of the map-making process in the error estimation. The details of the data reduction steps and procedure are described in a series of BISTRO papers \citep[e.g.,][]{wa17, kw18, wa19}. The POL-2 data reduction was done with a 4\arcsec\ pixel size, because larger pixel sizes can increase the uncertainty due to the map-making process. It is worth noting that JCMT observations use a chopping mode and high-pass filtering to remove background atmospheric emission. This also filters out extended source structures with scales $\gtrapprox$ 3\arcmin. This filtering does not affect our analysis because the physical size of G33.92+0.11 is about 3\arcmin, but helps to remove possible contamination from extended background and foreground emission.

The output Stokes I, Q, and U images were calibrated in units of mJy/beam, using a flux conversion factor (FCF) of 725 mJy/pW \citep{de13}, and binned to a pixel size of 12\arcsec\ to improve the sensitivity. The polarization fraction and orientations were calculated for each 12\arcsec\ pixel, which is close to the beam size of 14\arcsec, hence 
yielding nearly independent measurements. The typical rms noise of the final Stokes Q and U maps is $\sim$ 1.1 m\jyb\ at the center of the map. The Stokes I image has a higher intensity-dependent rms noise, 1--5 m\jyb\ near the center of the map, as a result of the uncertainties due to the map-making process.

We debiased the polarization fraction with the asymptotic estimator \citep{wa74} as
\begin{equation}\label{eq:debias}
P=\frac{1}{I}\sqrt{(U^2+Q^2)-\frac{1}{2}(\sigma_{Q}^2+\sigma_{U}^2)},
\end{equation}
where $P$ is the debiased polarization percentage, and $I$, $Q$, $U$, $\sigma_{I}$, $\sigma_{Q}$, and $\sigma_{U}$ are the Stokes $I$, $Q$, $U$, and their uncertainties. The uncertainty of the polarization fraction was estimated using
\begin{equation}\label{eq:eP}
\sigma_{P}=\sqrt{\frac{(Q^2\sigma_{Q}^2+U^2\sigma_{U}^2)}{I^2(Q^2+U^2)} + \frac{\sigma_{I}^2(Q^2+U^2)}{I^4}}.
\end{equation}
The polarization position angle ($PA$) was calculated as
\begin{equation}\label{eq:PA}
PA=\frac{1}{2}\tan^{-1}(\frac{U}{Q}),
\end{equation}
and its corresponding uncertainty was estimated using
\begin{equation}\label{eq:ePA}
\sigma_{PA}=\frac{1}{2}\sqrt{\frac{(Q^2\sigma_{U}^2+U^2\sigma_{Q}^2)}{(Q^2+U^2)^2}} .
\end{equation}

\subsection{IRAM 30-m Observations}\label{sec:IRAM}
The \C18O (2-1) emission line can trace dense gas with a density of $\sim10^4$ cm$^{-3}$ \citep[e.g.,][]{ni15}, comparable to the typical densities of pc-scale filaments \citep[e.g.,][]{an14}.
Aiming at probing gas kinematics within filaments, 
\C18O (2-1) spectral line observations were performed with the IRAM 30-m telescope on November 20, 2011 using the on-the-fly mapping mode (project 179-11, PI:Hauyu Baobab Liu) toward G33.92+0.11 (R.A.=18$^{h}$52$^{m}$50.2$^{s}$ and Dec.=0\degr55\arcmin30\farcs0). 
The scan produced a 6\arcmin$\times$6\arcmin map, covering both the G33.92+0.11 hub and the surrounding filaments. The observations were carried out with the HERA receiver tuned at 219.755 GHz and the FTS spectrometer as the spectral backend with a spectral resolution of 50 kHz, 0.066 \kms, and a bandwidth of 625 MHz. At this frequency, the telescope delivered a beam size of 11.8\arcsec. The data were reduced with the standard procedure with the CLASS package of the GILDAS\footnote{available at http://www.iram.fr/IRAMFR/GILDAS} software \citep{pe05,gi13}. To improve the sensitivity, we binned the data to a spatial resolution of 14\arcsec. The resulting typical rms noise of the smoothed data was 0.2 K.

\section{Results}\label{sec:results}
\subsection{Dust Continuum}
Assuming that the observed polarization traces magnetically aligned dust grains \citep[e.g.,][]{cu82,hi88}, we rotate the polarization segments by 90\degr\ and show the resulting magnetic field orientations with a pixel size of 12\arcsec, overlaid on the Stokes $I$ map with a pixel size of 4\arcsec\ in \autoref{fig:pmap}. The total intensity image clearly shows a bright central hub at the center of G33.92+0.11, surrounded by faint extended structures. In addition to the major central hub, a second minor hub (hereafter western hub) can be seen west of the central hub. This western hub is connected to the central hub through two bridging filaments.
Contours are additionally used in \autoref{fig:pmap} 
to emphasize connecting structures. 
In particular the lower contours around the central hub are not simply circular, but they reveal extensions to the south and west, likely indicating
the presence of filaments connecting to the central hub. 

\subsection{Polarization Data Selection}
In order to ensure significant detections, we limit
polarization segments to $I/\sigma_{I}>10$ and
$P/\sigma_{P} >2$ within the field of \autoref{fig:pmap}. 
The selected 152 polarization measurements include 91 segments with $P/\sigma_{P} >3$ and 61 segments with $3>P/\sigma_{P} >2$. The $P/\sigma_{P} >3$ sample has a maximum $\sigma_{PA}$ of 9.3\degr\ with a mean $\sigma_{PA}$ of 5.7\degr. The $3>P/\sigma_{P} >2$ sample has a maximum $\sigma_{PA}$ of 13.0\degr\ with a mean $\sigma_{PA}$ of 10.7\degr. The distribution of the polarization fraction and how the polarization fraction is related to the total intensity is reported in \autoref{sec:app_pp}, where we examine whether the observed polarization traces magnetically aligned dust grains.

\subsection{Magnetic Field Morphology}
The observed magnetic field morphology shows an overall converging feature: the field is prevailingly north-to-south  (PA$\sim$0\degr) on the northern and the south-eastern side of the central hub, and it is east-west (PA$\sim$90\degr) in the west with some segments to the east that appear to be bent and turning to the center. Around the western hub, the field is along a northwest-southeast orientation (PA$\sim$120\degr) in the west, and it is more northeast-southwest (PA$\sim$40\degr) on the eastern side, becoming almost parallel to the bridging filaments connecting to the central hub.
Additionally, the changes in field orientations are
mostly smooth,
as the local magnetic field often displays an orientation similar to adjacent segments. 
Since the polarization segments are plotted per independent beam (14\arcsec\ = 0.5 pc), adjacent segments are mostly uncorrelated, and thus the observed features indicate that the magnetic field morphology indeed varies smoothly over parsec-scale.


\subsection{Gas Kinematics}
\begin{figure*}
\includegraphics[width=\textwidth]{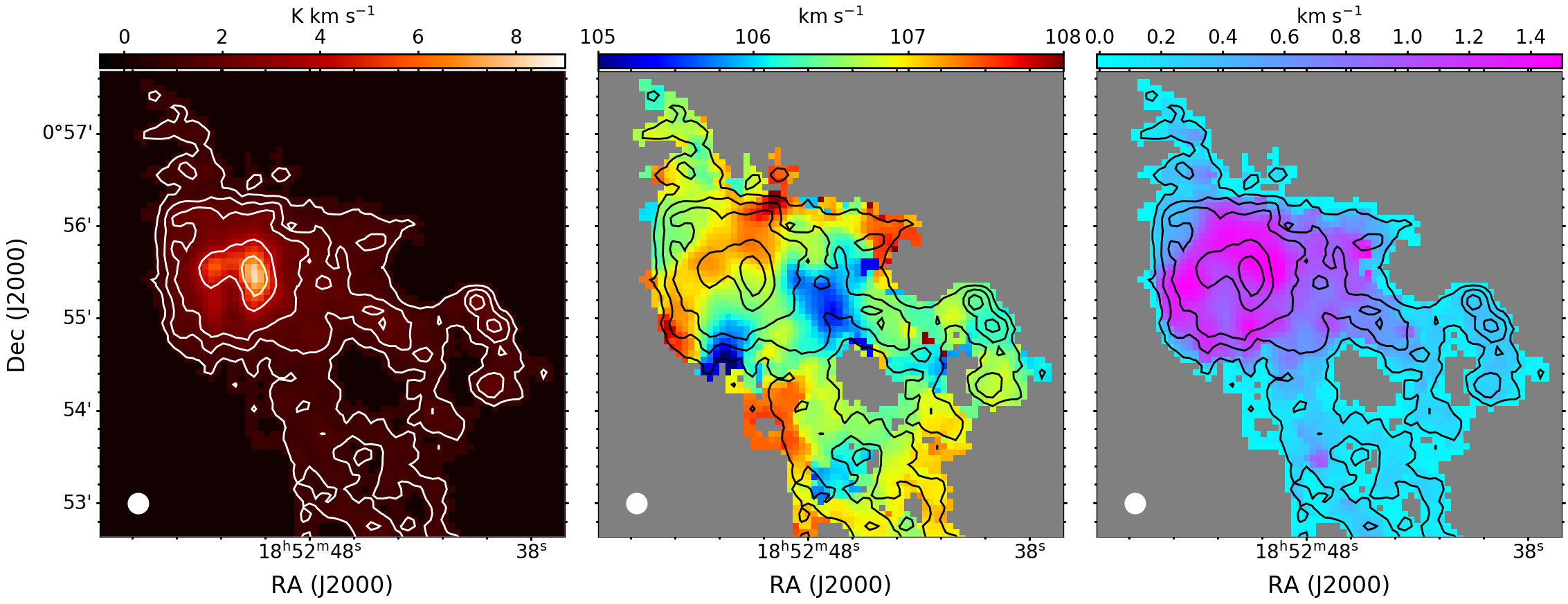}
\caption{(a) IRAM 30-m \C18O (2-1) integrated intensity map, smoothed to a resolution of 14\arcsec. The white contours represent integrated intensities of 1, 1.5, 2, 4, and 5 K~\kms\ with an rms noise level of $\sim$0.2 K. Clearly seen are two fragmented cores within the central hub, overlapping with the extended peak intensity region in the dust continuum map. (b) \C18O (2-1) centroid velocity map with integrated intensity contours. The velocity map reveals band-like patterns, along which the gas velocity only varies slightly. (c) \C18O (2-1) velocity dispersion ($\sigma_v$) map. The highest velocity dispersion regions coincide with the intensity peak, which is possibly related to the ongoing fragmentation as discussed in \autoref{sec:VGVD}}\label{fig:co_map}
\end{figure*}

In order to reveal the gas motion within G33.92+0.11, we performed Gaussian fits of the \C18O spectra pixel by pixel, using the python package PySpecKit \citep{gi11}, to estimate the gas centroid velocity and velocity dispersion. The pixels with a signal-to-noise ratio (SNR) $<$3 in the data cube were excluded from the fit. As most of the spectra show only one dominant component, we used a single-Gaussian model to trace the major gas motion. The \C18O integrated intensity, centroid velocity, and the velocity dispersion maps are shown in \autoref{fig:co_map}. We note that a few spectra in the hub center show possible multiple components. 
However, this is not a concern for our work here because we are 
focusing on the connecting extended structures 
towards the hub, and it is not our goal to resolve
the more complicated velocity structure inside
the central hub region. 

The integrated intensity map (left panel in \autoref{fig:co_map}) reveals a similar hub-filament morphology as seen in the dust continuum. Additionally, it resolves two fragmented cores within the central hub, which is visible but less obvious in the continuum image. Nevertheless, some of the faint extended continuum structures are not visible in the \C18O image due to its lower SNR. The \C18O centroid velocity map (middle panel in \autoref{fig:co_map}) presents several band-like structures where the LOS velocities are roughly similar. For example, a band with a LOS velocity of 107--108 km~s$^{-1}$ (yellow to red) extends from the hub center to the north-west and to the south-east. Another band with a LOS velocity of $\sim$106.5 km~s$^{-1}$ (green) extends from the center to the south and to the north-east. 
The derived \C18O velocity dispersion (right panel in \autoref{fig:co_map}) clearly increases from outside to inside, and the locations of the peak velocity dispersions match the central hub and some local intensity peaks. This peaks in dispersion can possibly be caused by gravitational acceleration. 

In summary, the similar features and extensions seen in the IRAM 30-m line and the JCMT 850 $\mu$m continuum data together with their very 
good spatially overlapping coverage suggest that the 
two data sets are tracing the same underlying structures. This forms our basis to further investigate how local gas motions correlate with the magnetic field.


\begin{figure*}
\includegraphics[width=\textwidth]{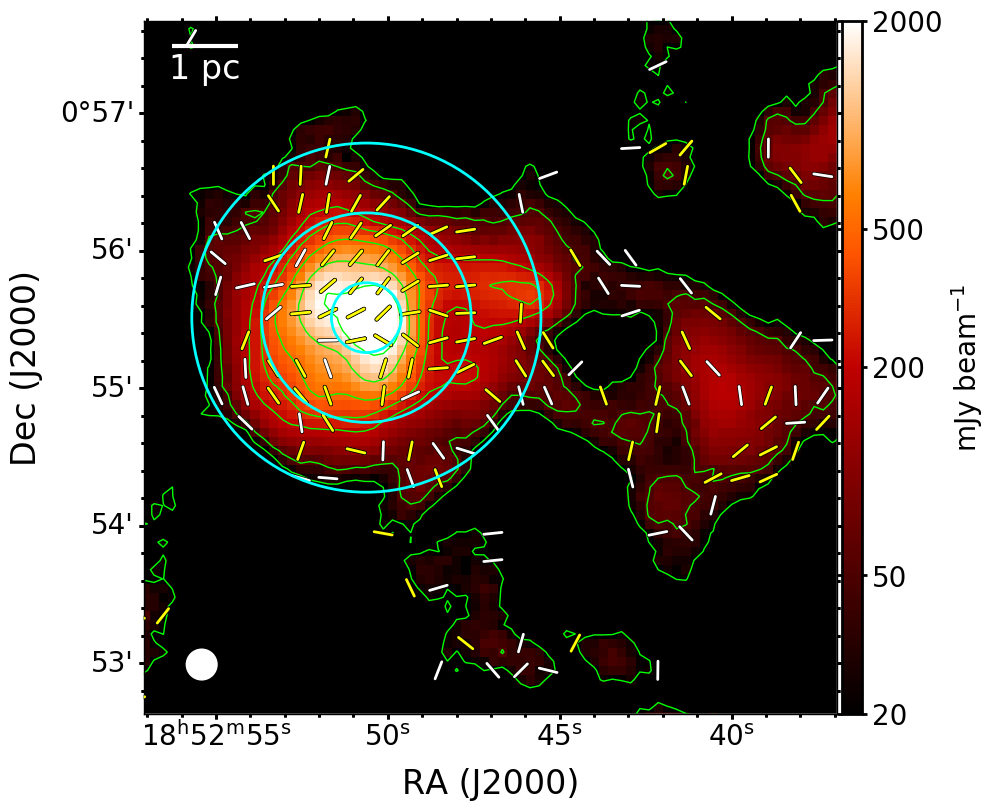}
\caption{B-field orientations (segments) sampled on a 12\arcsec\ grid overlaid on 850 $\mu$m dust continuum (color and contours), sampled on a 4\arcsec\ grid, of the G33.92+0.11 region. The segments are selected with the criteria $I/\sigma_{I}>10$ and $P/\sigma_{P}>2$, and rotated by 90\degr\
to represent magnetic field orientations. The yellow and white segments display the larger than 3$\sigma$ and 2--3$\sigma$ polarization detections. The green contours show the total intensity at 20, 50, 200, 300, 500, 1000, and 2000 m\jyb. The white circle in the bottom left corner is the JCMT beam size of 14\arcsec. The white scale bar in the upper left corner denotes 1 pc. The cyan circles label the 1, 3, and 5 FWHM (30.5\arcsec) areas, where we perform the magnetic field strength estimate in \autoref{sec:BS}. The Stokes I rms noise is $\sim$1--5 m\jyb\,, depending on pixel intensities, within the central 3\arcmin\ area.}\label{fig:pmap}
\end{figure*}


\section{Analysis}\label{sec:ana}
In the following sections, we aim at extracting the spatial properties of the filamentary structures, the local gravitational force, and the local velocity gradient. Together with the observed magnetic field morphology, these properties are used to investigate how these physical quantities possibly correlate with each other {\it locally}. 
The local aspect is central here because we are 
aiming at studying the detailed interplay between
the various parameters which is not possible 
from an analysis of global properties alone.

\subsection{Filament Identification}\label{sec:id_F}
In order to identify the ridges of filamentary structures 
in the JCMT 850 $\mu$m Stokes I image
we adopt the $DisPerSE$ algorithm \citep{so11}. 
This algorithm is designed to identify topological features within an image based on the Morse Theory. $DisPerSE$ first identifies the critical points (maxima, minima, and saddle points) in an image. Filaments are then identified as a series of multiple maxima-saddle point pairs connected by segments tangent to the intensity gradient field. In order to filter out insignificant structures, an intensity contrast threshold, named persistence threshold, is applied to exclude those maxima-saddle point pairs where the intensity differences are less than the threshold. The resolution of the identified filaments is roughly equal to that of the input image (4\arcsec), which is adequate to sample our beam size (14\arcsec).

\begin{figure*}
\includegraphics[width=\textwidth]{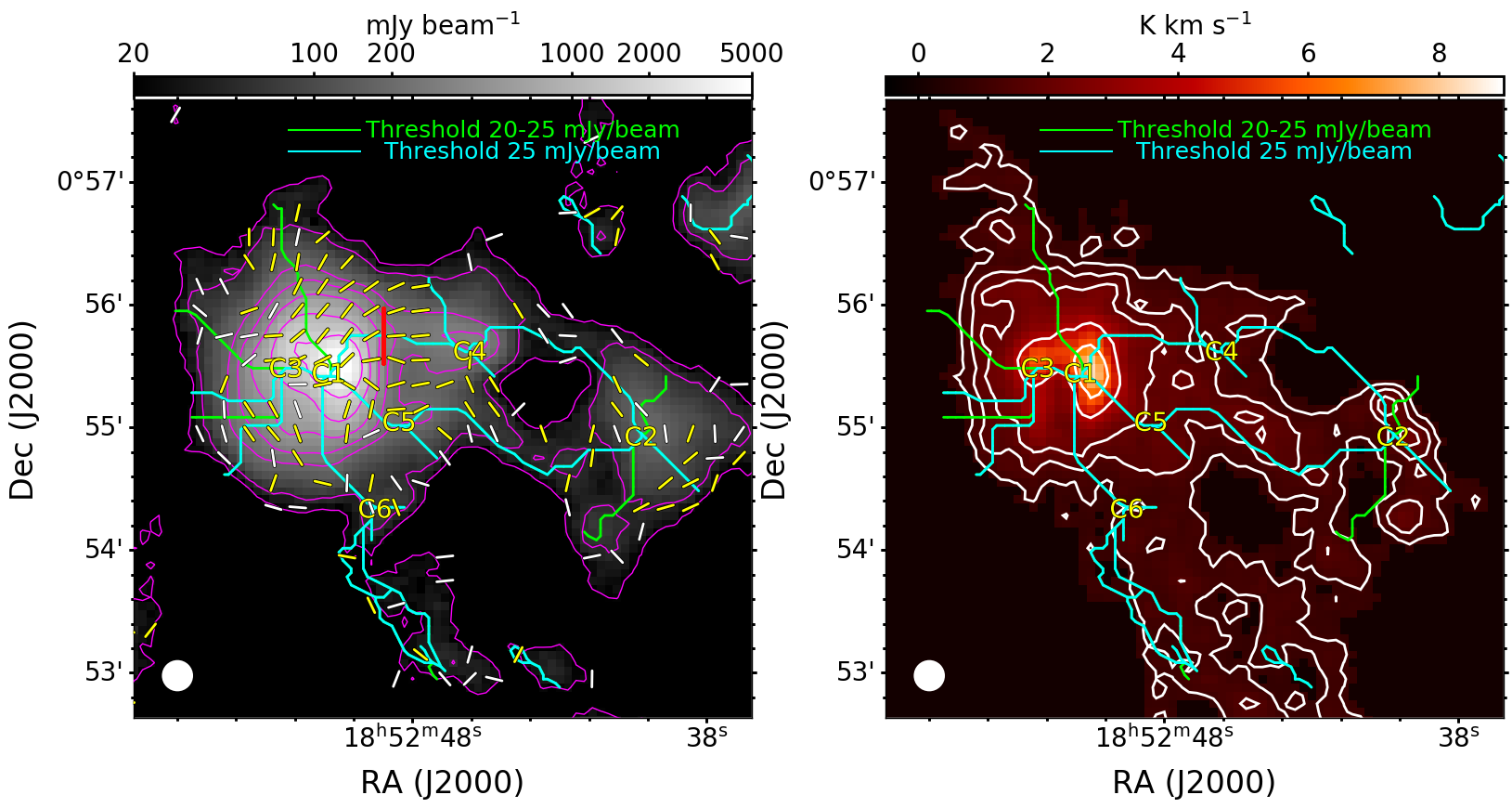}
\caption{(a) Filaments identified in the 850 $\mu$m dust continuum map using the $DisPerSE$ algorithm. 
The cyan filaments are identified with a threshold of $>$25 mJy/beam. Lowering the threshold to 20--25 mJy/beam leads to the additional filaments in green.
The magenta contours show the 850 $\mu$m intensities of 20, 50, 200, 300, 500, 1000, and 2000 mJy/beam. C1 and C2 label the major filament converging points in the central and western hub. C3--C6 are minor converging points. The red segment illustrates a 9-pixel ($\sim$1.1 pc) cut perpendicular to the filament with a radial intensity profile shown in \autoref{fig:fcut}. (b) Filaments identified as in panel (a), overlaid on the \C18O (2-1) integrated intensity map. The white contours denote integrated intensities of 1, 1.5, 2, 4, and 5 K~\kms.}\label{fig:fi}
\end{figure*}

\autoref{fig:fi}(a) shows the identified filaments with thresholds of 20 and 25 m\jyb, respectively. These thresholds correspond to a level of 4 and 5$\sigma$, where $\sigma$ is the maximum rms noise of 5 m\jyb\ near the image center. The filamentary networks identified with the two thresholds both present converging features. The longer filaments all connect to the major converging points (C1 and C2). The shorter filaments either connect to C1 and C2,  or merge into the longer filaments at the minor converging points (C3--C6). The major converging points coincide with the positions of the central and western hub, and the minor converging points match the local intensity peaks. 

\autoref{fig:fi}(b) further displays the identified filaments overlaid on the \C18O integrated intensity map. As the \C18O integrated intensity map reveals a morphology similar to the 850 $\mu$m continuum map, the identified filaments also roughly trace the filamentary structure in the \C18O intensity map. The converging points C1 and C3 match the locations of the two \C18O intensity peaks, revealing two fragmented cores,  within the central hub. Nevertheless, since the SNR in the \C18O integrated intensity map is much lower than in the 850 $\mu$m continuum map, some of the fainter filaments are less apparent. As a consequence, our further analysis is based on the filaments identified from the 850 $\mu$m continuum map. 
Moreover, we will mainly use the filaments identified with the higher persistence threshold of 25 m\jyb\
in order to identify high-SNR features. 
The possible bias due to the selected threshold
is discussed in \autoref{sec:full_cor}.

\begin{figure}
\includegraphics[width=\columnwidth]{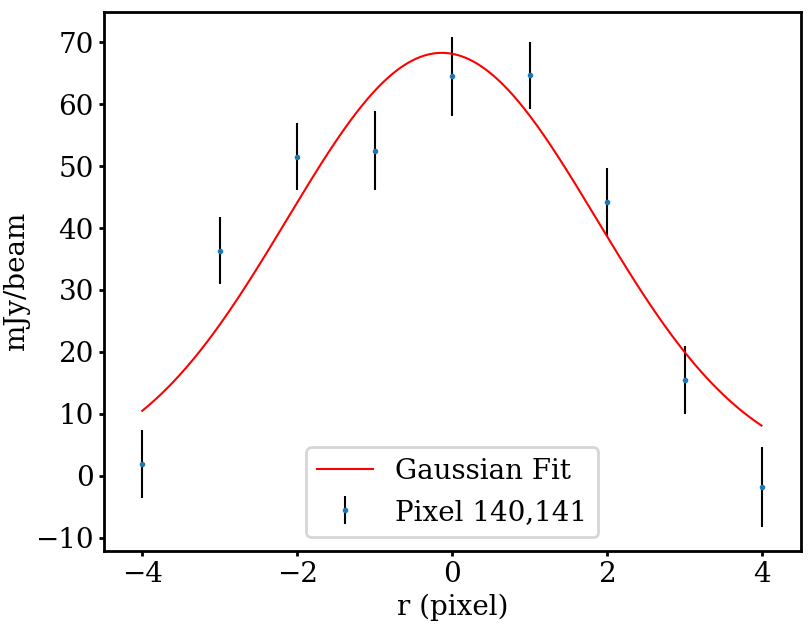}
\caption{Example of filament radial intensity profile along the red cut shown in \autoref{fig:fi}. The red line shows its best-fit Gaussian. Measurement uncertainties are displayed with vertical error bars.}\label{fig:fcut}
\end{figure}

In order to estimate the filament width, we use the python package $FilChap$\footnote{
Bugs in the Dec 2018 version of $FilChap$ were corrected prior to its application to the work here.}
\citep{su19}.
Since the intensity of the identified filaments often varies significantly, e.g., from tens of m\jyb\ in the faint envelope to thousands of m\jyb\ in the central hub, we do not expect the filament widths to be constant along the filaments. Hence, we fit individual radial intensity profiles extracted at each pixel position along the filament ridge. We apply a bootstrap method to fit these radial profiles with a Gaussian function. To estimate the fitting uncertainties, we use a Monte Carlo approach to generate 100 simulated profiles based on the observed intensities and uncertainties. The uncertainties of the fitting parameters are then estimated from the standard deviation of the fitting results to the 100 simulated samples. Fits with a width smaller than three times the uncertainties are excluded from the further analysis. The fitted Gaussian widths ($\sigma_F$) are deconvolved and converted to the FWHM ($\Delta F_{deconv}$) via $\Delta F_{deconv}^2= 8ln2\sigma_F^2-\theta^2_{beam}$, where $\theta_{beam}$ is the JCMT beam size (0.48 pc). An example of the radial intensity profile with its best fit is shown in \autoref{fig:fcut}. 

\begin{figure}
\includegraphics[width=\columnwidth]{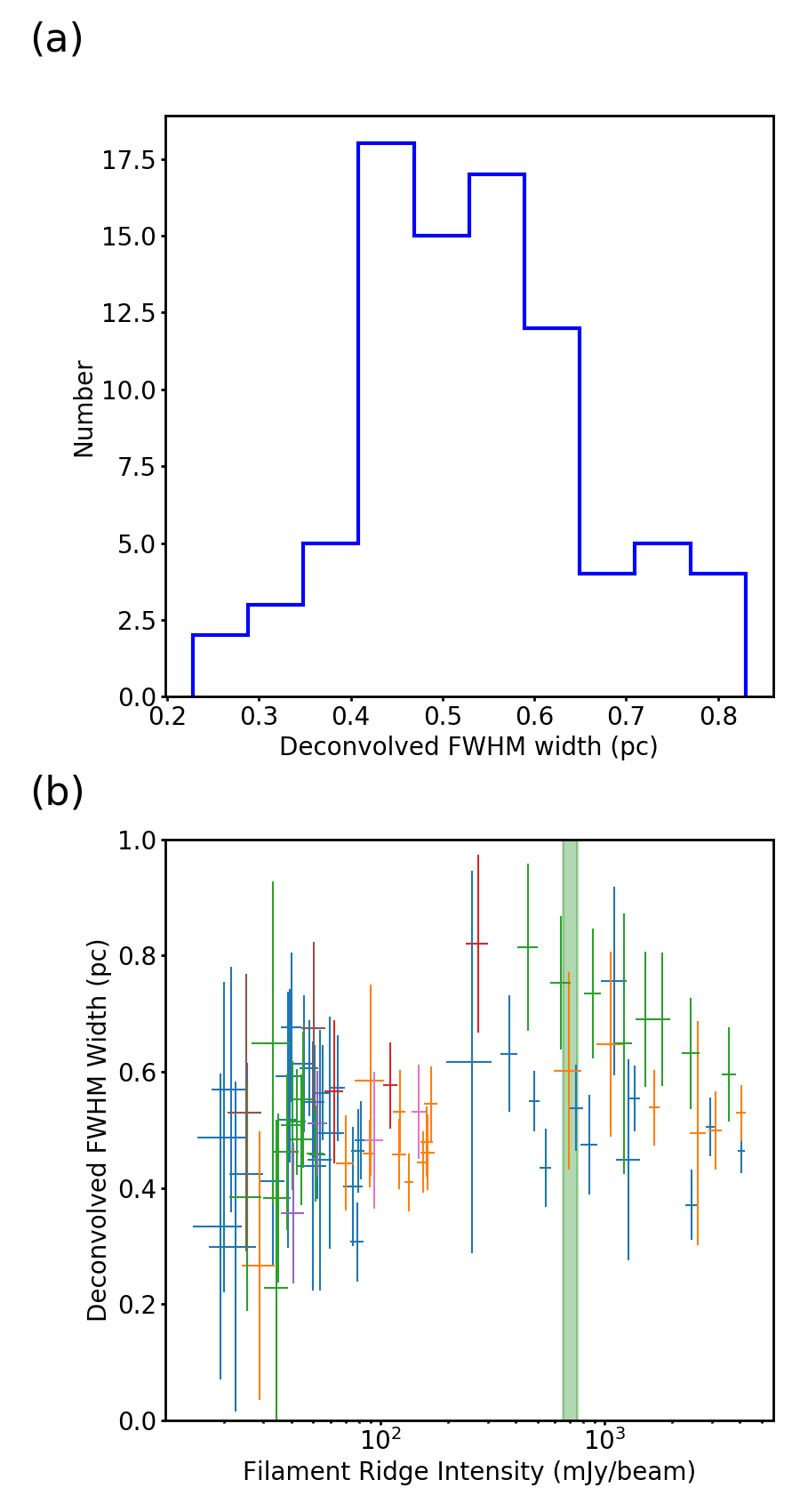}
\caption{(a) Histogram of deconvolved filament widths. The widths are estimated from the radial intensity profiles at each pixel along the identified filament ridges, as illustrated in \autoref{fig:fcut}. (b) Deconvolved filament widths vs. ridge intensities. Widths extracted along the same filament are shown with identical color. The filament width is not a constant. The mean width increases from 0.5 to 0.7 pc as the peak intensities increase from 200 to 1000 m\jyb, and decreases again as the intensities increase beyond 1000 m\jyb. The green area labels the critical linear density accounting for the support of thermal, non-thermal, and magnetic field pressure (see \autoref{sec:fpro}).}\label{fig:fi_width}
\end{figure}

\autoref{fig:fi_width}(a) shows the histogram of all deconvolved filament widths, mostly ranging from 0.4 to 0.6 pc. This is comparable to our spatial resolution of 0.48 pc. Thus, our data can only marginally resolve the filaments, and consequently are insufficient to probe the internal structure within the filaments. \autoref{fig:fi_width}(b) illustrates how the filament widths vary with the ridge intensities. The mean filament widths grow from 0.5 to 0.7 pc as the ridge intensities increase from $<$200 to $\sim$500-1000 m\jyb, and they decrease again to $\sim$0.5 pc as the ridge intensities grow beyond 2000 m\jyb. 

\subsection{Local Gravitational Field}\label{sec:gfield}
In order to investigate whether gravity influences the formation of the converging filaments, we estimate the projected gravitational vector field from the JCMT 850 $\mu$m continuum data. Since G33.92+0.11 is a distant object, foreground dust might bias the measured column density. We have tried to estimate the temperature and column density from the $Herschel$ 250, 350, and 500 $\mu$m continuum data, but found that the measured column density is generally too high ($\sim10^{22}$ cm$^{-2}$) in the diffuse ($I_{850}<100$ m\jyb) areas. In contrast, with the large-scale extended emission being filtered out by the JCMT scan mode, foreground contamination in the JCMT data is also minimized. The column density in the diffuse areas estimated from the JCMT 850 $\mu$m continuum data, adopting a constant temperature of 20 K from $Herschel$ data, is consistent with the typical value of $\sim10^{20}$--10$^{21}$ cm$^{-2}$. 
Hence, we are using the JCMT 850 $\mu$m continuum data to derive the column density of our target.


Following the development of the polarization-intensity gradient technique in \citet{koch12a,koch12b}, 
the local projected gravitational force acting at a pixel position ($\vec{F_{G,i}}$) can be expressed as the vector sum of all gravitational pulls generated from all surrounding pixel positions as
\begin{equation}
\vec{F_{G,i}} = kI_i\sum_{j=1}^{n}\frac{I_j}{r_{i,j}^2}\hat{r},
\end{equation}
where $k$ is a factor accounting for the gravitational constant and conversion from emission to total column density. $I_i$ and $I_j$ are the intensity at the pixel position $i$ and $j$, $n$ is the total number of pixels within the area of relevant gravitational influence. $r_{i,j}$ is the plane-of-sky projected distance between the pixel $i$ and $j$, and $\hat{r}$ is the corresponding unity vector. 
The above equation assumes a constant conversion between dust and total column density.
This leads to a {\it local gravitational vector field}, 
specifying a {\it local direction} and a {\it local magnitude}
of the gravitational pull at every selected pixel.
For our analysis here, 
we only focus on the directions of the local gravitational forces. Since we are not using 
the absolute magnitude, knowing the spatial
distribution of dust is sufficient, assuming that its distribution is a fair approximation for the 
distribution of the total mass. 
In the above equation we thus assume k=1.
When calculating the local gravitational field,
a lower threshold for the surrounding diffuse
and large-scale emission is introduced below which a gravitational influence is neglected. 
This is justified because this emission is 
weak and additionally less important because of 
an increasing $1/r^2$ suppression in the above equation, and furthermore, this diffuse emission tends to be rather symmetrical which means that any small gravitational pulls largely
cancel out.
Based on this framework, the local role of gravity and magnetic field, their relative importance, spatial variations and systematic features, and statistical properties are interpreted and analyzed across a sample of 50 star-forming regions in \citet{koch13,koch14}.

\begin{figure}
\includegraphics[width=\columnwidth]{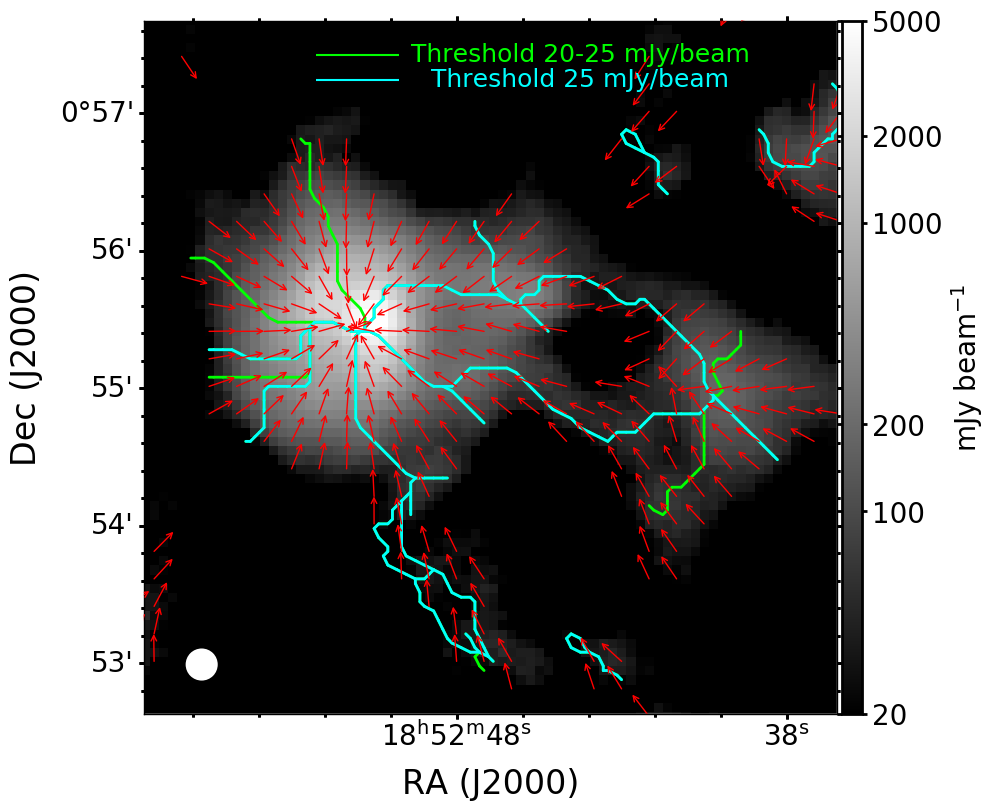}
\caption{Projected local gravitational field (red arrows). 
The gravitational force vectors are displayed with uniform lengths to emphasize their directions. The green and cyan lines label the identified filaments
as introduced in \autoref{fig:fi}(a).
The projected gravitational field is converging to the central hub, similar to the identified filaments. }\label{fig:gmap}
\end{figure}

For our analysis we mask the pixels with an 850 $\mu$m intensity lower than 20 m\jyb.
\autoref{fig:gmap} displays the local gravitational vector field. In most of the areas, the local gravity is pointing toward the central hub. Around the western hub, the gravitational field is pointing toward the western hub on the western side, but on the eastern side, it is turning around to
point to the more massive central hub. This morphology indicates that the massive central hub dominates the overall gravitational field in this area.

\subsection{Local Velocity Gradient}\label{sec:vg}
\begin{figure}
\includegraphics[width=\columnwidth]{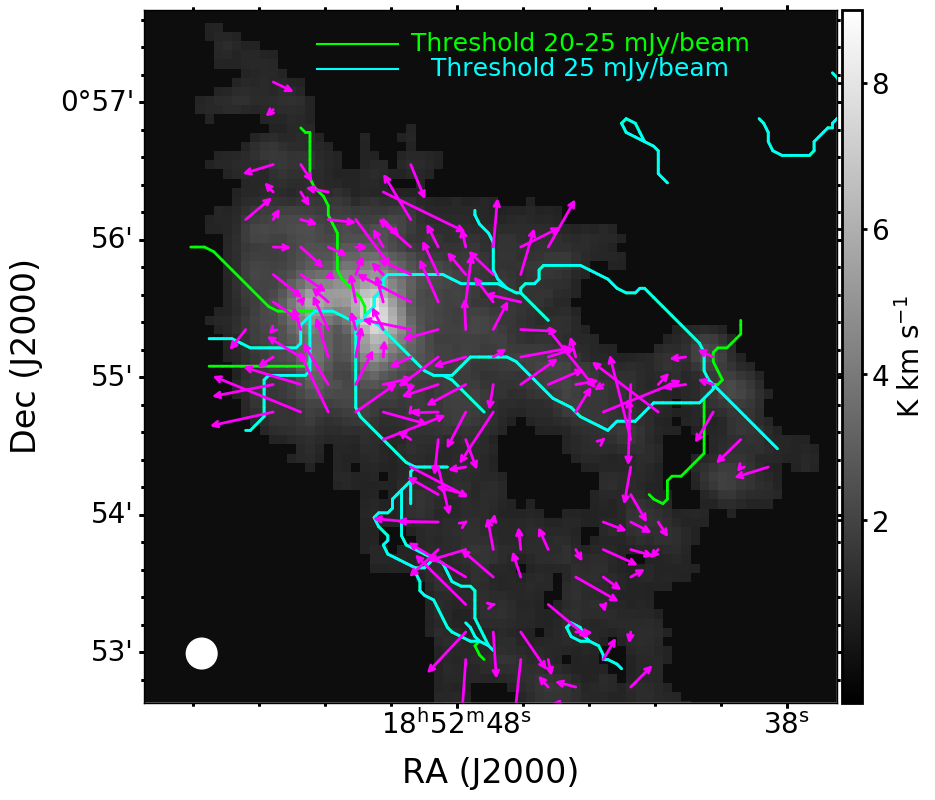}
\caption{Local LOS velocity gradient field (magenta arrows) overlaid on the \C18O (2-1) integrated intensity map. The magnitude is proportional to the 
length of the arrows. 
The green and cyan lines are the identified filaments as introduced in \autoref{fig:fi}(a). In several locations, the directions of the local velocity gradients seem to be perpendicular to the filaments in the outer areas, but become more disordered nearing the central hub. The magnitude of the velocity gradient appears to decrease as the distance to the central hub decreases.}\label{fig:vgmap}
\end{figure}

In order to investigate whether the gas kinematics possibly correlate with the filamentary structures, we calculate velocity gradients from the \C18O centroid LOS velocity map (\autoref{fig:co_map}(b)). In order to probe the local gas motion, we calculate {\it local} velocity gradients pixel by pixel, instead of only obtaining a global velocity gradient over an entire filament. To avoid calculating gradients from correlated pixels, the local velocity gradient at a pixel along the ridge of a filament is derived from the centroid velocities of the four nearest neighbouring independent beams (separated by 3 pixels, 12\arcsec). The resulting local velocity gradient field is plotted in \autoref{fig:vgmap}.

This velocity gradient field is less organized than the orientations of the filaments and the gravitational force field.
It seems to be perpendicular to the filaments in the diffuser areas, but then changing to become more random near the central hub. The magnitude of the local velocity gradient appears to decrease as the distance to the central hub decreases. However, a more thorough statistical analysis is necessary to understand whether these trends are significant or not (see later in \autoref{sec:VGVD}).

\subsection{Correlations and Trends among Filaments, Magnetic Field, Gravity, and Gas Kinematics}\label{sec:4par}
\subsubsection{Overall Statistical Correlations}\label{sec:4par_all}
To investigate how the physical parameters interact in the HFS, we perform systematic statistical analyses to reveal possible correlations among the orientations/directions of:
filaments (F), magnetic fields (B), gravitational force (G), and gas velocity gradient (VG). 

In order to estimate the filament orientations pixel by pixel, we select the identified filaments within 5 pc of the central hub. The filaments made up of fewer than 6 pixels ($\approx$2 independent beams) are excluded because they are not well resolved. For the $ith$ pixel along the ridge of a filament, we fit the positions of the $(i-2)th$ to the $(i+2)th$ consecutive pixels along the filament with a straight line to estimate the local filament orientation.

\begin{figure*}
\includegraphics[width=\textwidth]{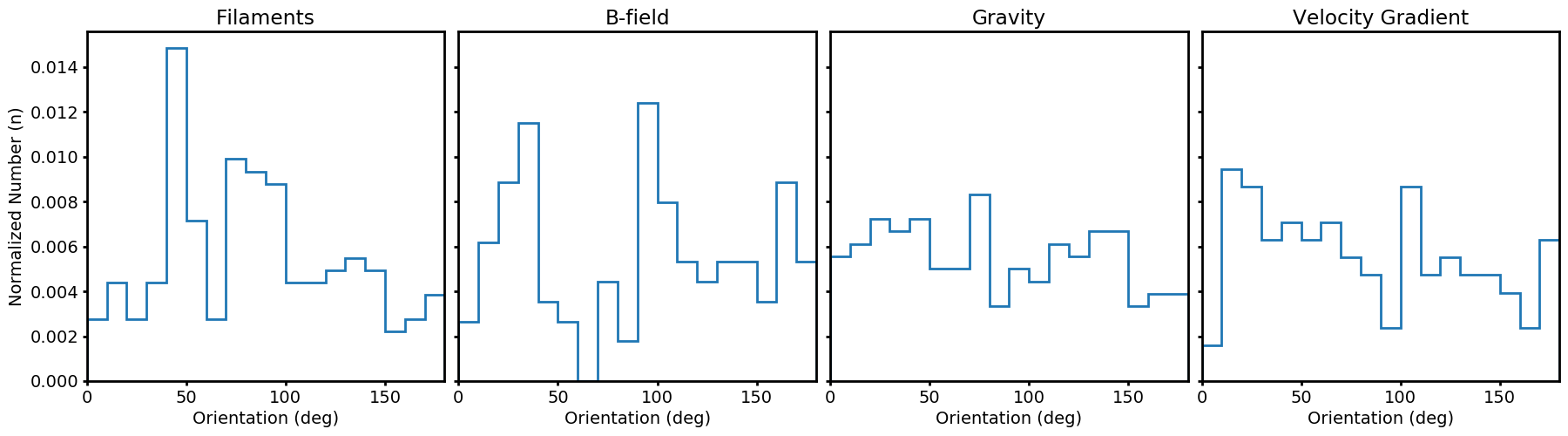}
\caption{Normalized histograms of local orientations of filaments, magnetic field, gravity, and velocity gradients. The position angle is defined following the IAU standard, PA=0\degr\ referring to north and increasing counterclockwise to east. }\label{fig:hist_4p}
\end{figure*}

We select the filament orientations, the magnetic field orientations, the gravitational force vectors, and the velocity gradients within 5 pc of the central hub as our sample to investigate their possible correlations.
\autoref{fig:hist_4p} presents the histograms of the local orientations of filaments, magnetic field, gravity, and velocity gradients. Given the limited number of filaments, the resulting histogram shows two main peaks corresponding to the longest filaments extending to the west and southwest. The other three histograms display broad distributions with possibly again two broad though less pronounced peaks.

An all-pairwise comparison of the four parameters 
(F, B, G, VG)
is performed.
Parameters are spatially matched as following. 
For each measurement of one parameter, we select a nearest measurement of another parameter within a radius of 18\arcsec (1.5 beam). These two measurements are then defined as one associated pair for the two parameters. 
A differential orientation ($\Delta PA$) is calculated for each associated pair. The resulting six distributions of all pairwise $\Delta PA$ are presented in \autoref{fig:hist_dpa}. 

A Kolmogorov–Smirnov (KS) test is used to examine whether the $\Delta PA$ distributions differ from random distributions. For the KS test, we use a random distribution as the null hypothesis, and estimate the probability (p-value) of the random distribution to match the observed distribution. A threshold of p=0.05 (95\% confidence interval) is used to reject the null hypothesis and identify the presence of a significant non-random trend. A significant non-random trend is further classified as parallel-like or perpendicular-like by whether the observed cumulative distribution is above or below the random distribution (diagonal line). The cumulative distributions of all six pairwise $\Delta PA$ are plotted in \autoref{fig:hist_dpa_ks}. Among these pairs, the most significant correlation ($p<0.001$) is found between local gravity and filament orientations
(G vs F). This is expected because both the filaments and local gravity appear to overall converge towards the central hub, and consequently, the $\Delta PA$ distribution is clearly peaking at small values, and hence parallel-like. In addition, the orientations between local velocity gradient and magnetic field (VG vs B) are significantly non-random with $p=0.004$. This 
$\Delta PA$ distribution reveals a trend of being perpendicular-like, i.e., the local velocity gradient being perpendicular to the local magnetic field.

We note that a p-value higher than 0.05 only indicates that the observed $\Delta PA$ distribution is compatible with a random distribution. This can be caused by either an underlying randomly distributed $\Delta PA$ or a small sample size. Since our sample sizes in the denser area are often small, high p-values found for this area should not yet be 
taken as a final answer to not identifying any
systematic structure in the data (see later section
\autoref{sec:4par_int}). 
Moreover, a KS test is not able to capture any 
possible spatial trends, as discussed in the
following section. 

\subsubsection{Spatial Trends along Filaments}\label{sec:4par_spatial}
\begin{figure}
\includegraphics[width=\columnwidth]{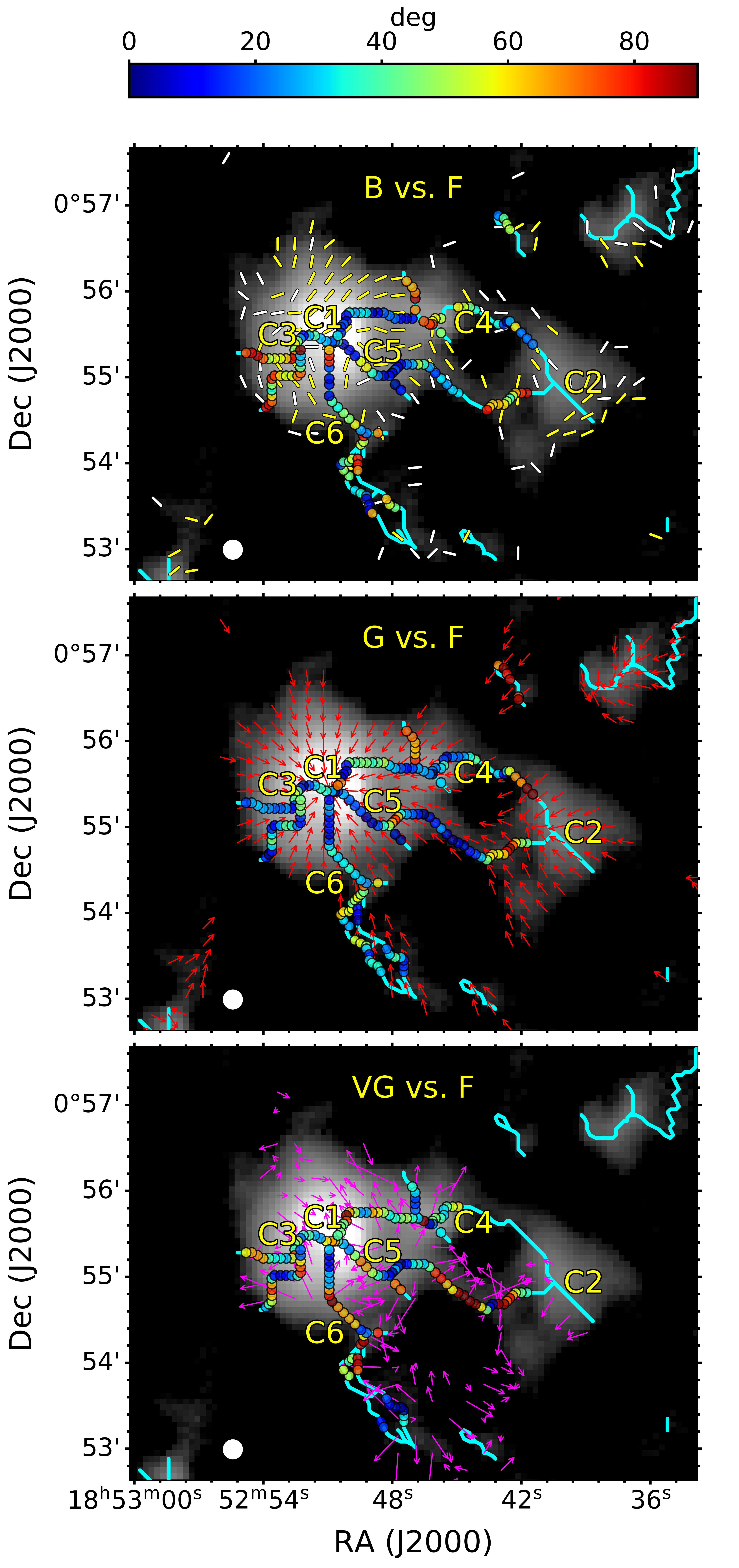}
\caption{Differential orientation maps for 
filament vs magnetic field, gravity, and velocity gradient (from top to bottom), 
overlaid on the 850 $\mu$m intensity. The cyan lines are the filaments identified in \autoref{fig:fi}. The yellow and white segments represent the magnetic field orientations from \autoref{fig:pmap}, the red arrows are the projected local gravity, and the magenta arrows show the local velocity gradients. 
Filled color-coded circles (color wedge) are the pairwise
differential orientations $\Delta PA$.
}\label{fig:dpa_map}
\end{figure}

Histograms cannot visualize possible spatial trends
among parameters. Therefore,  
we also show the $\Delta PA$ maps for the B-F, G-F, and VG-F pairs in \autoref{fig:dpa_map}. Maps for all pairs are in \autoref{sec:full_cor}. These maps reveal that these $\Delta PA$ may vary along the filaments, possibly because the roles of these physical parameters change. Despite the overall statistical randomness between magnetic field and filament orientations (\autoref{fig:hist_dpa_ks}), the B-F $\Delta PA$ map shows that the magnetic field appears to be more parallel to the filaments near the center of the dominating central hub, but occasionally becomes perpendicular to the short filaments connecting to the minor converging points C2, C3, and C4. This suggests that these short filaments might interact with the longer filaments differently. The G-F $\Delta PA$ map displays an alignment between gravity and filaments almost everywhere. There is non-alignment (local gravity perpendicular to filament) only in a few short filaments near the minor converging point C4. Although the overall $\Delta PA$ of VG-F is random (\autoref{fig:hist_dpa_ks}), the VG-F $\Delta PA$ map indicates that the local velocity gradients tend to be more perpendicular to the filaments in the outskirts of this system, and then become more random near the hub center. This may hint that the gas kinematics near the hub center become more complicated, turbulent, or chaotic.

\subsubsection{Trends with Intensity}\label{sec:4par_int}
\begin{figure*}
\includegraphics[width=\textwidth]{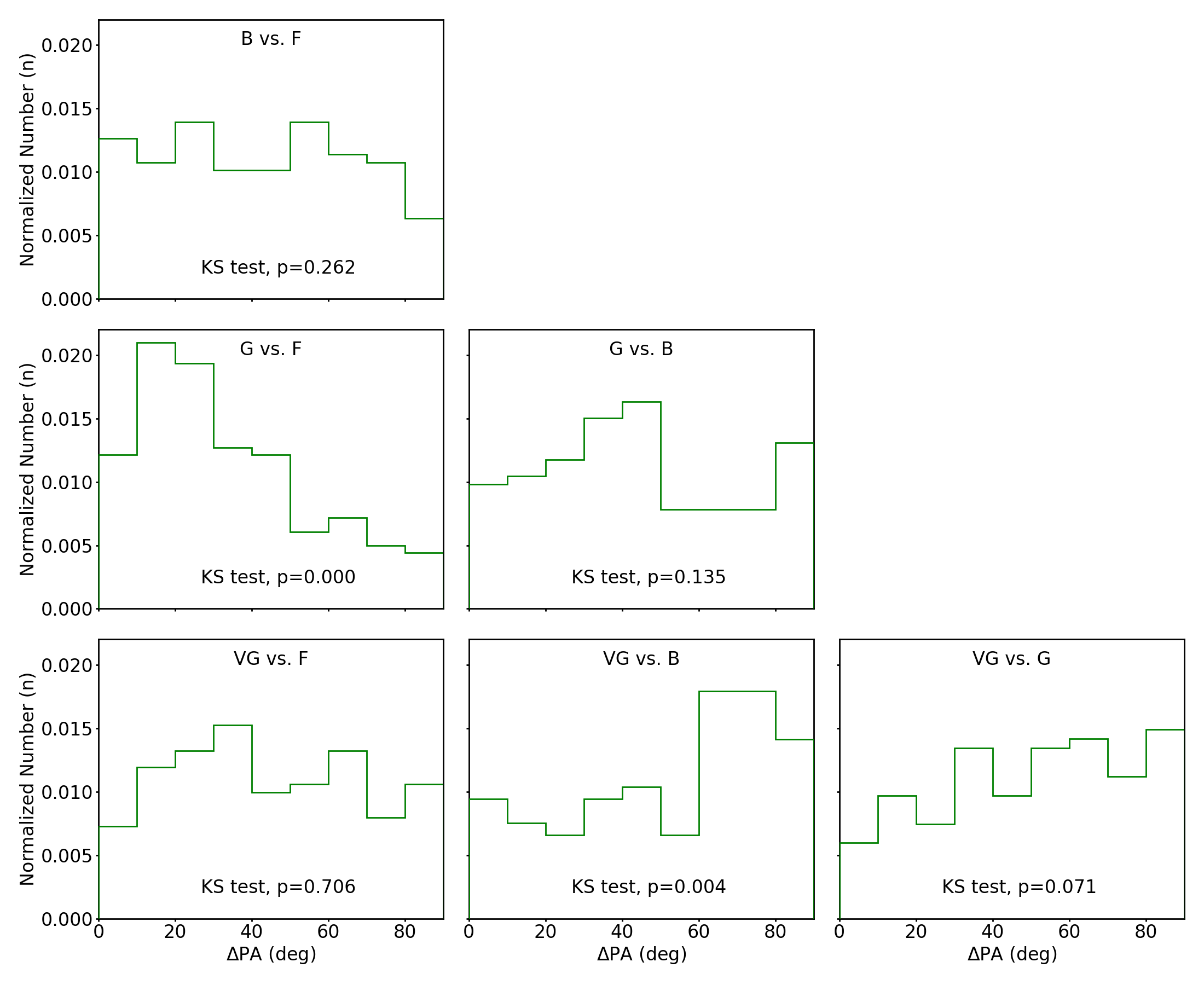}
\caption{Normalized histograms of local pair-wise  differential orientations among filaments, magnetic field, gravity, and velocity gradients. p-values from KS tests are given in each panel.}\label{fig:hist_dpa}
\end{figure*}

\begin{figure*}
\includegraphics[width=\textwidth]{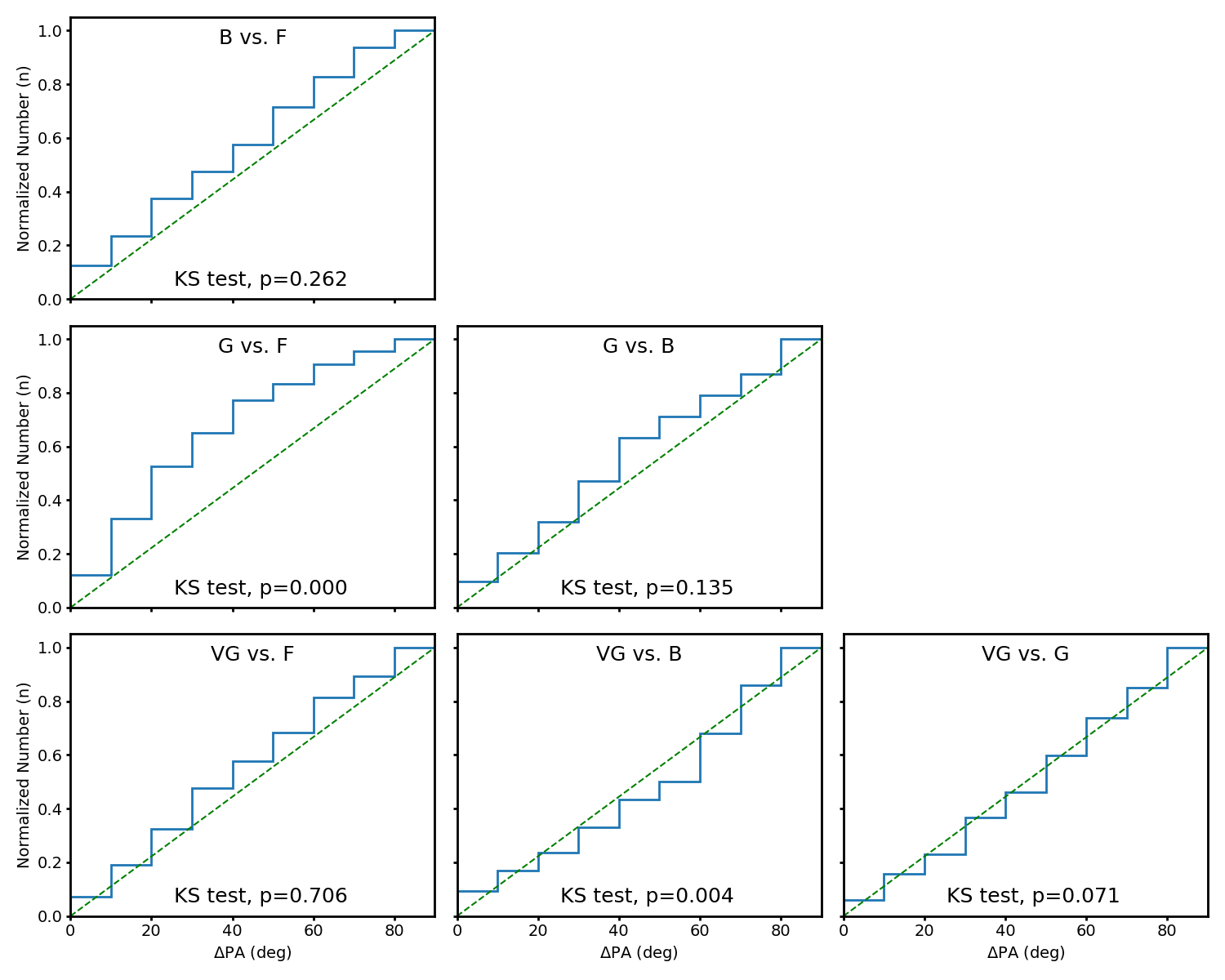}
\caption{Cumulative distributions of differential orientations among the associated filaments, magnetic field, gravity, and velocity gradients, based on 
\autoref{fig:hist_dpa}. The green dashed diagonal line indicates the random distribution. The resulting p-values of the KS-tests are given in each panel.
The G vs F distribution can be classified as parallel-like while the VG vs B distribution is 
perpendicular-like.
A threshold of p$<$0.05 is adopted as a non-random distribution with a 95\% confidence interval.}\label{fig:hist_dpa_ks}
\end{figure*}

\begin{figure*}
\includegraphics[width=\textwidth]{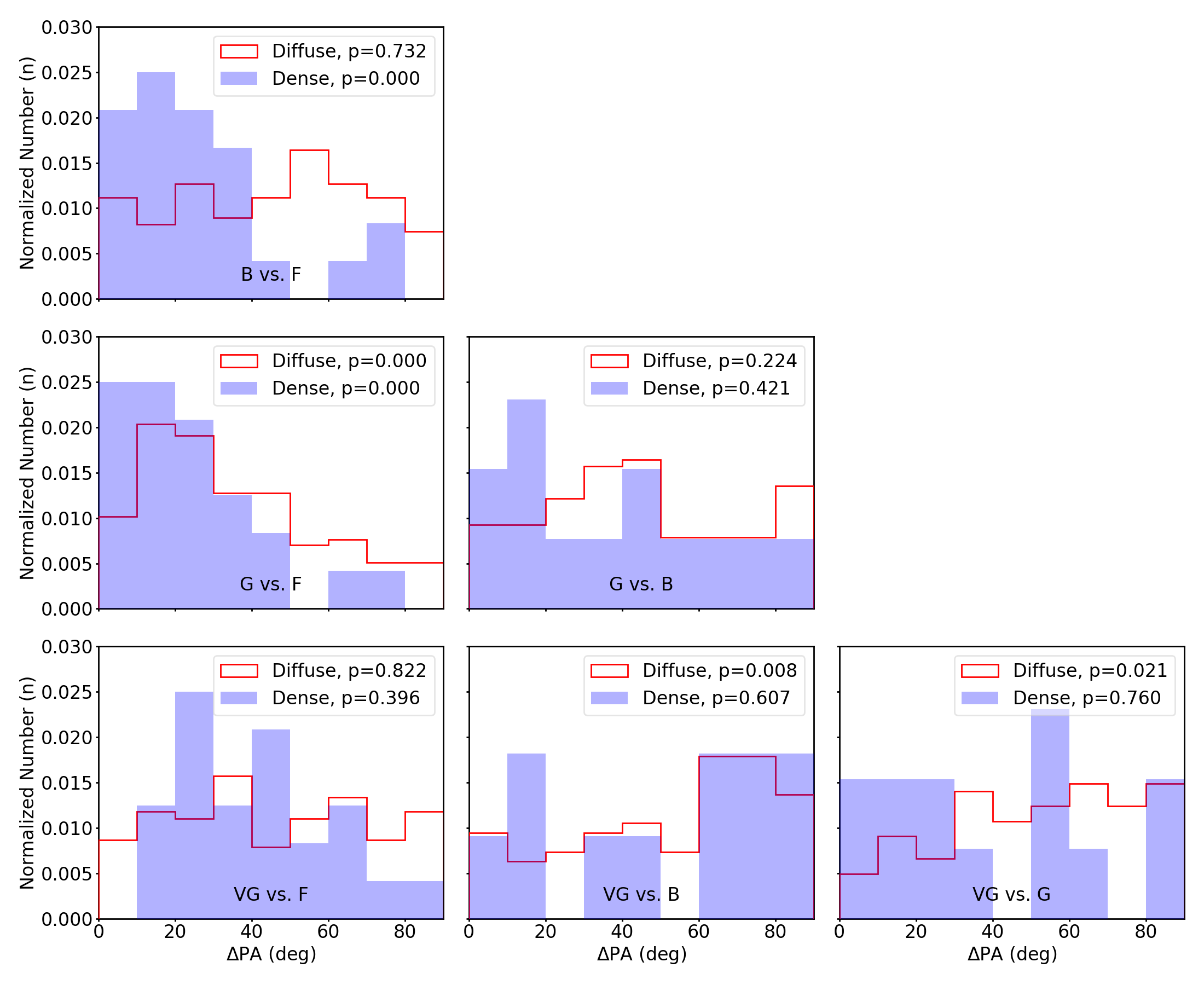}
\caption{Normalized histograms of the pairwise differential orientations between filaments, magnetic field, gravity, and velocity gradients, 
for two different intensity ranges.
The blue and red histograms represent the $\Delta PA$ selected in the dense ($I_{850}>$ 1000 m\jyb) and the diffuse ($I_{850}<$ 1000 m\jyb) areas. The KS-test results against the random distribution are listed for the two areas.}\label{fig:hist_dpa_2I}
\end{figure*}

Given the spatial trends identified in the previous
section, we now further investigate a possible evolution from the outer area to the hub center.
In order to do so, 
we separate the matched pairs in \autoref{fig:hist_dpa} into a dense-area group ($I>1000$ m\jyb) and a diffuse-area group ($I<1000$ m\jyb). This boundary is chosen so that filaments in the dense-area group are likely supercritical (see \autoref{sec:fpro} and \autoref{fig:fi_width}).
The histograms of $\Delta PA$ for the two groups are shown in \autoref{fig:hist_dpa_2I},
and the distributions of $\Delta PA$ versus the local intensity is reported in \autoref{sec:full_cor}.
Since the samples in the diffuse area are larger than in the dense area, the overall statistics in \autoref{fig:hist_dpa} are dominated by the diffuse samples, and thus, identical trends are found in the isolated diffuse samples in \autoref{fig:hist_dpa_2I}. Nevertheless, from the comparison between the low- and high-intensity groups, we find that the interaction between the physical parameters evolves with the local density: The B-F pair evolves from no tendency, i.e., no prevailing alignment (p=0.732) in the low-intensity regions to aligned (p$<$0.001) in the high intensity regions. In contrast, the VG-B and VG-G pairs change from prevailingly perpendicular (p=0.008 and p=0.021) in the low-intensity regions to no tendency (p=0.607 and p= 0.760) in the high-intensity regions. 
{\it In summary, these findings reveal a tightening alignment between filaments, gravity, and magnetic field with growing intensity, while the local velocity structures become more disordered.}


\subsection{Global Stability of G33.92+0.11}\label{sec:BS}
In this section, we aim to investigate the global stability of G33.92+0.11 by evaluating the balance between gravity, magnetic field, and gas kinematics. In order to estimate the magnetic field strength from the polarization data, we use both the Davis-Chandrasekhar-Fermi (DCF) method \citep{da51,ch53} in \autoref{sec:DCF}, and the structure function (SF) method \citep{ho09} in \autoref{sec:SF}. The calculated magnetic energy scale is compared with the gravitational and kinematic energies using the virial theorem in \autoref{sec:virial}.

\subsubsection{Davis-Chandrasekhar-Fermi Method}\label{sec:DCF}
The DCF method assumes that kinematic and magnetic energy are in equipartition.
Therefore, the level of magnetic field perturbation
(traced by the magnetic field angular dispersion $\delta \phi$)
that can result from turbulence
(traced by the LOS non-thermal velocity dispersion $\sigma_{v,NT}$)
is determined from both the plane-of-sky magnetic field strength ($B_{pos}$) and the gas volume density $\rho$ by
\begin{equation}\label{eq:CF}
B_{pos}=Q~\sqrt[]{4\pi \rho}\frac{\sigma_{v,NT}}{\delta \phi},
\end{equation}
where Q is a factor accounting for complex magnetic field and inhomogeneous density structures. \citet{os01} suggested that $Q=0.5$ yields a good estimation of the magnetic field strength on the plane of sky if the magnetic field angular dispersion is less than 25\degr .

As the role of the magnetic field may evolve with  local density, we separate G33.92+0.11 into three areas with diameters of 1, 3, and 5 FWHM size of the central hub (30.5\arcsec) as shown in \autoref{fig:pmap}. We select the polarization segments in each area and calculate its polarization angular dispersion. 
In order to remove a possible bias from an underlying large-scale magnetic field morphology contributing to the dispersion,  
we calculate differences in polarization angles only using nearest pixel pairs. 
The polarization angular dispersion is derived as
\begin{equation}
\delta \phi = \sqrt{\frac{1}{N-1}\sum_{i=1}^{N}\delta PA_{i}^2},
\end{equation}
where $\delta PA$ is the absolute difference between two polarization angles for every possible nearest-polarization pair, N is the total number of pairs, and the factor $1/(N-1)$ is to debias the population standard deviation estimator. The calculated $\delta PA$ 
within the 1, 3, and 5 FWHM areas are listed in \autoref{tab:CF}.

To estimate the mean volume densities in the areas, we have first calculated a column density map using the 850 $\mu$m continuum data assuming a constant dust temperature of 20 K and a dust opacity $\kappa$ of 0.012 cm$^2$/g \citep{hi83}. The total mass of each area is integrated from the column density over the selected area. The mean volume density is then estimated from the total mass and assuming a spherical volume. 

In order to estimate the mean non-thermal velocity dispersion, we first average the observed \C18O line widths in each area. Assuming a gas kinematic temperature ($T_\mathrm{kin}$) of 20 K, the thermal velocity dispersion for \C18O is $\sqrt{\frac{k_B T_\mathrm{kin}}{m_{\mathrm{C}^{18}\mathrm{O}}}}=0.09\pm0.01$ km~s$^{-1}$. The thermal velocity dispersion is then removed from the observed line width to obtain the non-thermal velocity dispersion as
\begin{equation}\label{eq:vdisp}
\sigma^2_{v,NT}=\sigma^2_{obs} - \frac{k_B T_\mathrm{kin}}{m_{\mathrm{C}^{18}\mathrm{O}}}
\end{equation}
where $\sigma_{obs}$ is the observed C$^{18}$O Gaussian line width, and $m_{\mathrm{C}^{18}\mathrm{O}}$ is the molecular weight.
The resulting magnetic field strengths for the three areas are listed in \autoref{tab:CF}.

\subsubsection{Structure Function Method}\label{sec:SF}
\citet{ho09} expanded the DCF method, assuming that the magnetic field is composed of an ordered large-scale component ($B_0$) and a turbulent component ($B_t$). The ratio of these two components determines the polarization angular dispersion as
\begin{equation}
\delta \phi = \left[\frac{\langle B_t^2\rangle}{\langle B_0^2\rangle}\right]^{\frac{1}{2}},
\end{equation}
where $\langle...\rangle$ denotes an average. Hence, the DCF equation (\autoref{eq:CF}) can be rewritten 
as 
\begin{equation}\label{eq:SF_B}
B_{pos}=\sqrt[]{4\pi \rho}~\sigma_{v,NT} \left[\frac{\langle B_t^2\rangle}{\langle B_0^2\rangle}\right]^{-\frac{1}{2}},
\end{equation}
where the original correction factor $Q$ in the DCF equation is neglected, because the structure function approach is expected to remedy the possible bias originating from the depolarization effect due to the turbulent magnetic field component. \citet{ho09} further showed that the ratio of turbulent-to-magnetic energy can be estimated from the structure function using the following equation:
\begin{equation}\label{eq:SF}
1-\langle\cos[\Delta\Phi(\ell)]\rangle \simeq \frac{1}{N}\frac{\langle B_t^2\rangle}{\langle B_0^2\rangle}(1-e^{-\ell^2/2(\delta^2+2W^2)})+a\ell^2,
\end{equation}
where $\Delta \Phi (\ell)$ is the structure function that describes the difference between polarization angles separated by a distance $\ell$. $\delta$ and $a$ are unknown parameters, representing the turbulent correlation length and the first order Taylor expansion of the large-scale magnetic field structure. $W$ is the telescope beam radius, which is $6\farcs2$ at 850 $\mu$m for the JCMT. $N$ is the number of turbulent cells along the line of sight and within the telescope beam, and can be estimated from:
\begin{equation}\label{eq:ncell}
N=\Delta'\frac{\delta^2+2W^2}{\sqrt{2\pi}\delta^3},
\end{equation}
where $\Delta'$ is the effective cloud thickness. Fitting the above equation to the observed $1-\langle\cos[\Delta\Phi(\ell)]\rangle$ vs $\ell$ distribution yields the three unknown parameters $\delta$, $\frac{\langle B_t^2\rangle}{\langle B_0^2\rangle}$, and $a$.

\begin{figure}
\includegraphics[width=\columnwidth]{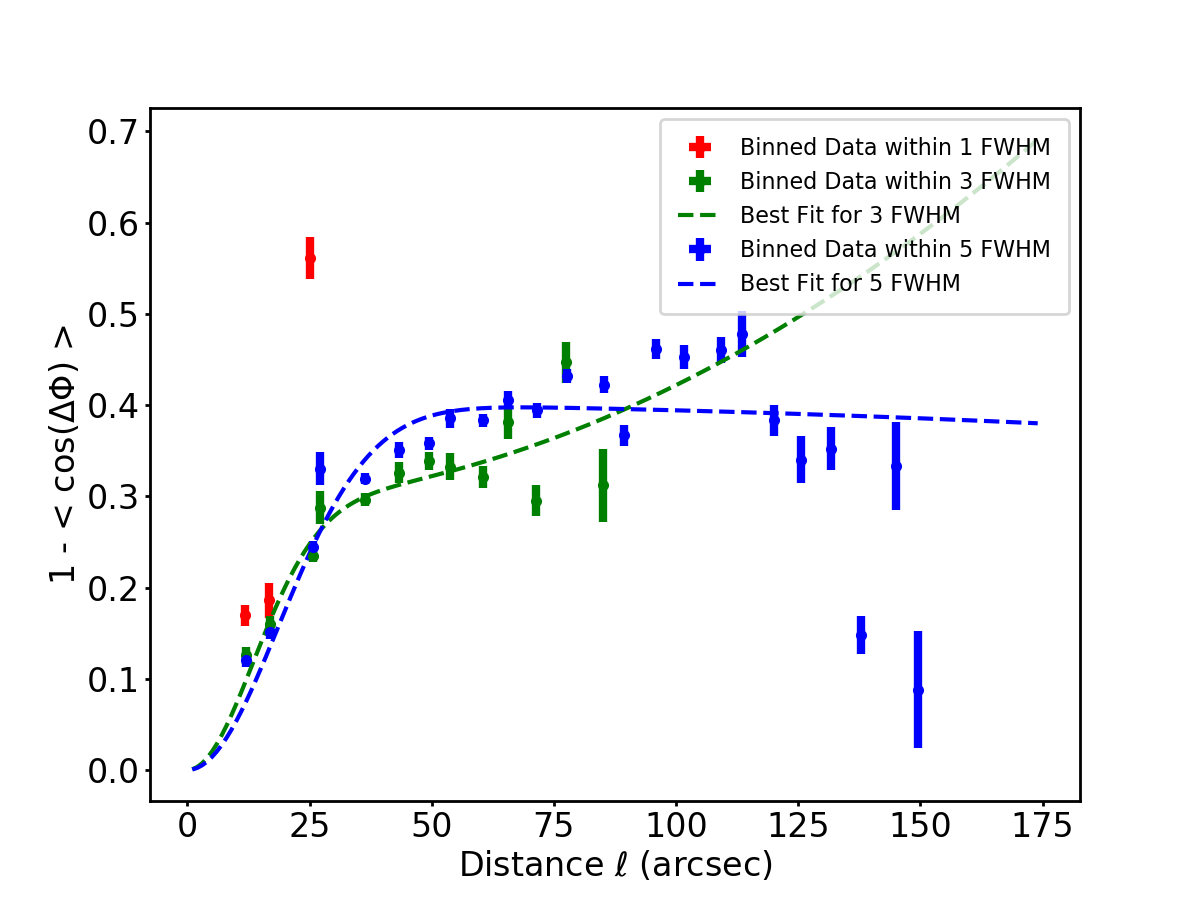}
\caption{Structure functions for polarization data selected within 1, 3, and 5 FWHM areas. The dashed lines show the best fits following \autoref{eq:SF}.}\label{fig:SF}
\end{figure}

Identical to the three areas defined for the DCF method, 
we calculate three separate dispersion functions. 
The derived $\cos[\Delta\Phi(\ell)]$ are averaged in bins of $12\arcsec$.
The resulting angular dispersion functions with their best fits of \autoref{eq:SF} are plotted in \autoref{fig:SF}. 
The fitting parameters are given in \autoref{tab:CF}. We note that no fit is attempted for the 1 FWHM area because this small area only yields three binned $\cos[\Delta\Phi(\ell)]$ values. Generally, the magnetic field strengths estimated from the DCF method are larger than the ones from the SF method
by a factor of $\sim$1.5. 
A possible reason for this discrepancy is that 
the observed $\delta\Phi$ are beyond the reliable range ($<25\degr$) for the DCF method as suggested by \citet{os01}.


To evaluate the relative importance between magnetic field and gravity, the mass-to-flux ratio ($\lambda_{obs}$) is calculated as
\begin{equation}
\lambda_{obs}=2\pi \sqrt{G}\frac{\mu m_{H}N_{H_2}}{B_{pos}},
\end{equation}
where $\mu$=2.33 is the mean molecular weight per H$_2$ molecule \citep{na78}. Due to the unknown inclination, the observed mass-to-flux ratio $\lambda_{obs}$ is only an upper limit. \citet{cr04} suggest that a statistical average factor of $1/3$ can be used to better estimate the mass-to-flux ratio accounting for the random inclinations in a sample of oblate spheroid cores, flattened perpendicular to the magnetic field. Thus, the corrected mass-to-flux ratio ($\lambda$) becomes 
\begin{equation}
\lambda=\frac{\lambda_{obs}}{3}.
\end{equation}
We note that this correction factor is determined from the cloud and magnetic field geometry. A correction factor of $\pi/4$ is suggested for a spherical cloud \citep{cr04}, and a factor of $3/4$ for a prolate spheroid elongated along the magnetic field \citep{pl16}. Using a correction factor of $1/3$, the estimated mass-to-flux ratios (2--3) are larger than unity for all areas. This suggests that this HFS is supercritical. The mass-to-flux ratios would be even more supercritical if other correction factors were used, and they would only become subcritical if the magnetic fields were nearly along the line-of-sight. A globally supercricial condition is in agreement with the overall converging filamentary structures aligned with gravity and the growing density along these filaments, because these features require dominating gravity driving mass to the center. 


\subsubsection{Virial Balance}\label{sec:virial}
The virial theorem is commonly used to analyze the balance between gravitational, magnetic, and kinematic energy in a molecular cloud. In Lagrangian form it can be written as
\begin{equation}\label{eq:virial}
\frac{1}{2}\ddot{I}=2(\mathcal{T}-\mathcal{T}_s)+\mathcal{M}+\mathcal{W},
\end{equation}
\citep[e.g.,][]{me56,mc07} where $I$ is a quantity proportional to the trace of the inertia tensor of the cloud. The sign of $\ddot{I}$ determines the acceleration of the expansion or contraction of the cloud. The term
\begin{equation}
\mathcal{T}=\frac{3}{2}M\sigma^2_{obs}
\end{equation}
is the total kinematic energy, where M is the total mass and $\sigma_{obs}$ is the observed total velocity dispersion. We neglect the surface kinetic term $\mathcal{T}_s$ because we aim at estimating the self-stability of an enclosed region. Nevertheless, we note that the presence of any external pressure could suppress gravitational support and enhance the cloud's instability. The magnetic energy term, without any force from an external magnetic field, is 
\begin{equation}
\mathcal{M}=\frac{1}{2}MV^2_A,
\end{equation}
where $V_A=B/\sqrt{4\pi\rho}$ is the Alfv\'{e}n velocity
and $\rho$ is the mean density. We note that the magnetic field morphology is not accounted for in this magnetic energy term. As the DCF and SF method only constrain the plane-of-sky magnetic field component, we use the statistical average to correct and estimate the total magnetic field strength as $B = (4/\pi)B_{pos}$ \citep{cr04}. The term
\begin{equation}
\mathcal{W}=-\frac{3}{5}\frac{GM^2}{R}
\end{equation}
is the gravitational potential of a sphere with a uniform density $\rho$ and a radius $R$. 

We derive the ratios of $\abs{\mathcal{T}/\mathcal{W}}$ and $\abs{\mathcal{M}/\mathcal{W}}$ in \autoref{tab:CF} for the three selected areas to evaluate the relative importance among kinematic, magnetic, and gravitational energy. The estimated energy scales generally indicate a relative importance of $\abs{\mathcal{W}}>\abs{\mathcal{T}}\gtrapprox\abs{\mathcal{M}}$, suggesting that gravity overall dominates the system. The kinematic energy is comparable to the magnetic energy, as $\abs{\mathcal{T}/\mathcal{W}}$ is only larger than $\abs{\mathcal{M}/\mathcal{W}}$ by a factor of 1.2--2.5. The relative importance of these three parameters is similar in the three selected areas, although the relative importance of kinematic energy seems to slightly increase near the central area.

As $\mathcal{T}$ and $\mathcal{M}$ are positive and $\mathcal{W}$ is negative, the sign of $\ddot{I}$ in \autoref{eq:virial} is positive (expanding) if $\abs{\frac{2\mathcal{T+M}}{\mathcal{W}}} > 1$ or negative (contracting) if $\abs{\frac{2\mathcal{T+M}}{\mathcal{W}}}<1$. All the derived values for  $\abs{\frac{2\mathcal{T+M}}{\mathcal{W}}}$ are lower than unity, suggesting that the kinematic and magnetic energy cannot support the system, and hence the whole system is contracting globally.

\begin{deluxetable*}{ccccccccccc}
\tablecaption{Parameters for Magnetic Field Strength Estimates and Virial Analysis.\label{tab:CF}}
\renewcommand{\thetable}{\arabic{table}}
\tablenum{1}
\tablehead{\colhead{Regions} & \colhead{$\frac{\langle B_t^2\rangle}{\langle B_0^2\rangle}$} & \colhead{$\delta$} & \colhead{$\sigma_{v,NT}$} & \colhead{$\delta \phi$\tablenotemark{b}} & \colhead{$n_{H_2}$} & \colhead{$B_{pos}$} & \colhead{$\lambda$} & \colhead{$\abs{\frac{\mathcal{T}}{\mathcal{W}}}$} & \colhead{$\abs{\frac{\mathcal{M}}{\mathcal{W}}}$} & \colhead{$\abs{\frac{2\mathcal{T+M}}{\mathcal{W}}}$}\\
\colhead{} & \colhead{} & \colhead{pc} & \colhead{(km~s$^{-1}$)} & \colhead{(deg)} & \colhead{(cm$^{-3}$)} & \colhead{($\mu$G)} & \colhead{} & \colhead{} & \colhead{} & \colhead{}}
\startdata
\multicolumn{8}{c}{DCF method}\\
\hline
1 FWHM sphere & 1.5$\pm$0.1\tablenotemark{a} & ... & 1.3$\pm$0.1 & 35.2$\pm$1.3 & $(1.0\pm0.1) \times 10^5$ & $230\pm8$ & $2.0\pm0.1$ & 0.17$\pm$0.01 & 0.08$\pm$0.01 & 0.42$\pm$0.01  \\
3 FWHM sphere & 1.1$\pm$0.1\tablenotemark{a} & ... & 1.1$\pm$0.1 & 30.0$\pm$1.2 & $(1.0\pm0.1) \times 10^4$ & $74\pm3$ & $1.8\pm0.1$ & 0.15$\pm$0.01 & 0.10$\pm$0.01 & 0.41$\pm$0.01 \\
5 FWHM sphere & 1.0$\pm$0.1\tablenotemark{a} & ... & 0.8$\pm$0.1 & 29.3$\pm$1.9 & $(2.5\pm0.3) \times 10^3$ & $27\pm2$ & $2.1\pm0.1$ & 0.10$\pm$0.01 & 0.08$\pm$0.01 & 0.29$\pm$0.01 \\
\hline
\multicolumn{8}{c}{SF method}\\
\hline
3 FWHM sphere & 1.85$\pm$0.63 & 0.35$\pm$0.06 & 1.1$\pm$0.1 & ... & $(1.0\pm0.1) \times 10^4$ & $57\pm10$ & $2.3\pm0.4$ & 0.15$\pm$0.01 & 0.06$\pm$0.01 & 0.36$\pm$0.01 \\
5 FWHM sphere & 1.90$\pm$0.3 & 0.57$\pm$0.06 & 0.8$\pm$0.1 & ... & $(2.5\pm0.3) \times 10^3$ & $20\pm2$ & $2.8\pm0.3$ & 0.10$\pm$0.01 & 0.04$\pm$0.01 & 0.25$\pm$0.01 \\
\enddata
\tablecomments{$\delta$, $\sigma_{v,NT}$, $\delta \phi$, $n_{H_2}$, $B_{pos}$, $\lambda$, $\mathcal{T}$, $\mathcal{M}$, and $\mathcal{W}$ are the turbulence length scale, non-thermal velocity dispersion, magnetic field angular dispersion, H$_2$ volume density, plane-of-sky magnetic field strength, mass-to-flux ratio, kinematic energy (both thermal and non-thermal), magnetic energy, and gravitational energy, respectively. }
\tablenotetext{a}{The ratio of turbulent-to-large-scale magnetic field components is estimated assuming $[\frac{B_{t}}{B_{0}}]^2 = \frac{\delta\phi}{Q}^2$.}
\tablenotetext{b}{The angular dispersion is estimated from pairs of polarization segment separated by 14\arcsec, which is the smallest physical scale we can probe.}
{\addtocounter{table}{-1}}
\end{deluxetable*}


\section{Discussion}\label{sec:discussion}

\subsection{Filament Properties}\label{sec:fpro}
$Herschel$ observations suggest that filamentary structures are ubiquitous in molecular clouds, and are likely progenitors of star formation. These filaments have widths within a narrow range around $\sim$0.1 pc, and tend to be co-linear in direction to the longer extents of their host clouds \citep{an14}. Using the same DisPerSe algorithm, we have identified numerous filaments within the G33.92+0.11 HFS. The identified filaments show an
overall converging configuration where all filaments are connecting to the massive central hub. This configuration is consistent with other known HFS, e.g., G10.6-0.4 \citep{liu12b}, W49 \citep{ga13}, B59 \citep{pe12} and SDC13 \citep{wi18}, but different from other filamentary systems with co-linear orientations, e.g., IC5146 \citep{ar11} and B211/213 \citep{pal13}. 

In addition to the overall configuration, we have measured the width and the peak intensity along the identified filaments. The derived median deconvolved FWHM is 0.5 pc, which is larger than the common filament width of 0.09$\pm$0.07 pc in the $Herschel$ samples by a factor of 5 \citep{ar11,an14,ar19}. We further find that the filament widths grow with the ridge intensity from 0.5--0.7 pc until the  intensity reaches $\sim1000$ m\jyb, and then turn to decrease to 0.5 pc as the intensity further increases.

The variation in filament widths has been seen in models studying the evolution of externally pressurized filaments \citep{fi12,he13a,he13b}. In these models, the internal gas motion is driven by the accretion onto filaments. Consequently, the turbulent pressure grows as the filaments gain mass. This is consistent with the observed increase in velocity dispersion with peak intensity (right panel in \autoref{fig:co_map}). If the internal turbulent pressure grows faster than  gravity and any external pressure, the excess gas pressure will expand the filaments until the gas pressure is balanced again by gravity. On the other hand, once sufficient mass has been accreted, the filament becomes supercritical and starts to  contract, causing a turnover and eventually leading to a decreasing radius. These mechanisms result in a peaked dependence on filament width and column density, and the peak will be around 0.5--2 times the critical linear density, depending on the filament's inclination angle.

In order to show that the observed variation in filament widths is consistent with the models described above, we have calculated the critical linear density as follows. Considering the support of both thermal and non-thermal gas motion\footnote{The term ``non-thermal motion'' refers to the motion traced by the observed non-thermal velocity dispersion, which includes turbulence but possibly also line-of-sight components of infall and rotational motion. In massive star-forming regions, gravitationally driven infall and rotational motion can be one of the major components in the observed non-thermal velocity dispersion
\citep[e.g.,][]{he09,tr20}.
The kinematic energy in the virial analysis is, therefore, also derived with both the thermal and non-thermal component.}, the critical linear density is $M_{line,critical}=2(c_s^2+\sigma_{v,NT}^2)/G$ \citep{fi20}. As the observed \C18O velocity dispersion is 0.05 \kms\ in the diffuse area, dominated by the thermal motion, the critical linear density is dominated by the $c_s$ term, which is $\sim$0.2 \kms\ for a mean molecular weight of 2.33 at 20 K. In contrast to that, the observed velocity dispersion is $\sim$1.2 \kms\ in the central hub, dominated by the non-thermal motion $\sigma_{v,NT}$ term. This results in an increase of the $M_{line,critical}$ from $\sim20M_{\sun}/pc$ in the diffuse area to $\sim800M_{\sun}/pc$ in the dense area. 

Additionally to the above estimated support, magnetic fields can play a role in stabilizing filaments. However, the exact magnetic support is determined by the morphology of the magnetic field within a filament, for which our current data do not
have sufficient resolution. The general virial analysis for magnetized filaments shows that poloidal-dominated fields help supporting filaments against gravity while toroidal-dominated fields destabilize filaments \citep{fi20}. The critical linear density of a magnetized filament is $M^{mag}_{line,vir}=M_{line,critical}\times(1-\mathcal{M}/\mathcal{W})^{-1}$, where $\mathcal{M}$ is the magnetic energy per unit length (positive for poloidal fields and negative for toroidal fields) and $\mathcal{W}$ is the gravitational energy per unit length. To estimate the upper and lower limit of $M^{mag}_{line,vir}$, we adopt our estimated $\abs{\frac{\mathcal{M}}{\mathcal{W}}}\sim0.08$ (\autoref{tab:CF}). $(1\pm\mathcal{M}/\mathcal{W})^{-1}$ then yields 0.9 (purely toroidal) and 1.1 (purely poloidal). Hence, the $M^{mag}_{line,vir}$ is estimated to be $\sim$700--900 $M_{\sun}/pc$, corresponding to a H$_2$ column density of 4--5$\times10^{22}$ cm$^{-2}$ for a filament width of 0.7 pc and an 850 $\mu$m intensity of $\sim$530-660 mJy/beam assuming a dust temperature of 20 K. 

This estimated critical linear density (highlighted in \autoref{fig:fi_width} as a green shaded region) is within the intensity range of $\sim$500-1000 m\jyb\ where the maximum filament widths occur.
{\it We, therefore, conclude that 
the observed trend and change in filament widths are consistent with the predictions of externally pressurized filament models. Due to the support from the non-thermal kinematic energy, most of the filaments in G33.92+0.11 remain subcritical until entering the massive center area with a radius of $\sim$1 pc} (third contour in \autoref{fig:pmap}). 
We speculate that this additional support from non-thermal energy is one of the reasons why these filaments can transfer a significant amount of mass toward the center hub before self-fragmenting and forming massive clusters.

The filaments within G33.92+0.11 may not be the same type of filaments as identified by $Herschel$ \citep[e.g.,][]{an10,ar11}. The statistical analysis based on 599 filaments from the $Herschel$ Gould Belt survey \citep{ar19} shows that most of the filaments have a central column density ranging from 5$\times10^{20}$ to 2$\times10^{22}$ $cm^{-2}$, and the central column densities within the same filament are typically comparable. In contrast, several filaments in G33.92+0.11 show a variation in peak intensity of one or even two orders of magnitude. In addition, the dense filaments in G33.92+0.11 can reach a density of $\sim10^{23}$ $cm^{-2}$. Such massive examples are rare in the $Herschel$ Gould Belt survey samples.  \citet{ar13} measured the velocity dispersion of the filaments identified by $Herschel$ in IC5146, Aquila, and Polaris, and found that these filaments have total velocity dispersions of 0.2--0.6 \kms, which is less than half of the peak velocity dispersion seen in G33.92+0.11. As the strong turbulent pressure can significantly increase the critical linear density for the filaments in G33.92+0.11 -- by a factor of $\sim$40 compared to the $Herschel$ filaments -- the filaments in G33.92+0.11 are able to accrete and transfer a large amount of mass to the center before self-fragmenting and forming massive clusters.

\subsection{Origin of Local LOS Velocity Gradient}\label{sec:VG}
The statistics in \autoref{sec:4par} show that local gravity tends to be parallel to filaments (\autoref{fig:hist_dpa_2I}). Since the observed velocity gradients tend to be perpendicular to local gravity in the diffuse areas (\autoref{fig:hist_dpa_2I}), one might expect the local velocity gradients to also be perpendicular to filaments.
The bottom panel in \autoref{fig:dpa_map} illustrates that the local velocity gradients 
in the diffuse areas
tend to be perpendicular to the long major filaments that connect to the center hub (C1),
except for the regions near the minor converging points (C3, C4, and C5). The minor filaments that connect to the minor converging points appear to be aligned with the local velocity gradients. Thus, 
the filament-velocity gradient statistics
are biased by a mixture of major and minor filaments. Based on the map displaying spatial tendencies, we conclude that local velocity gradients in the diffuse areas are still significantly tied to filaments: they are perpendicular to major filaments, but parallel to minor filaments.

The above trend is similar to the filament-striation configuration in the Taurus B211/213 system. The $^{12}$CO and $^{13}$CO line observations in \citet{go08} reveal numerous fine and diffuse filamentary structures, named striations, perpendicular to the major filamentary cloud but parallel to local magnetic fields. \citet{pal13} further describe a significant velocity gradient parallel to the striations but perpendicular to the major filament. \citet{sh19} model these velocity structures and suggest that the striations represent mass accretion from the ambient cloud to the major filaments along the magnetic field. This magnetic-field-guided mass accumulation has also been seen in other filamentary systems, such as Serpens South \citep{fe14,ch20}, Musca \citep{co16} and G34.43+00.24 \citep{ta19}.

One major difference between G33.92+0.11 and the filament-striation systems is that the local velocity gradients in G33.92+0.11 tend to be perpendicular to, instead of following the magnetic fields (\autoref{fig:hist_dpa_2I}). Magnetic fields parallel to striations but perpendicular to major filaments are expected to be a feature of systems with strong magnetic fields \citep[e.g.,][]{su87,na08}. In such systems, magnetic fields regulate the collapse of a cloud, and thus guide the accretion flow toward major filaments \citep{na08}. However, a magnetic-field-regulated collapse is not the only mechanism that can produce accretion flows perpendicular to a filament. Simulations of filaments produced by colliding turbulent flows can also explain the observed local velocity gradients perpendicular to filaments because the material surrounding the filaments tends to collapse parallel to the shock-compressed layers \citep{go11,go14,ch20}. In this scenario, minor converging points are intersections of multiple accretion flows, where they merge and are redirected toward a major converging point. This might explain the observed alignment between filaments and local velocity gradients around the minor converging points because here the gas kinematics are dominated by filaments merging due to local gravity instead of accumulating surrounding ambient mass. This scenario seems to better explain the observed trend of local velocity gradients being perpendicular to local magnetic fields. Moreover, this scenario is also supported by the estimated relatively weak magnetic energy. 

The spatial resolution of our polarization observations can only marginally resolve the filaments. Hence, we cannot completely exclude the possibility that the magnetic field outside the filaments is actually more parallel to the velocity gradient, but the observation is dominated by the polarization inside the filaments, where the magnetic field becomes parallel to the filament axis. Nevertheless, such a turn-over magnetic field morphology is also a feature predicted by turbulent compression models \citep{pa01b,ch14,ch15,go18}.
This is because the magnetic fields perpendicular to filaments on large scales, might be either bent by shocked gas layers or pinched by gas motions along filaments, and thus become parallel to filaments near a ridge. This has recently been seen in NGC6334 \citep{ar20}. Hence, even when considering the possibly unresolved turn-over magnetic field morphology, the observed magnetic fields  perpendicular to local velocity gradients still favor a turbulent compression scenario. Finally, we note that our analysis is only based on the LOS velocity component, and a significant plane-of-sky velocity component might change this picture.

\subsection{Velocity Gradient Magnitude and Velocity Dispersion}\label{sec:VGVD}
\begin{figure}
\includegraphics[width=\columnwidth]{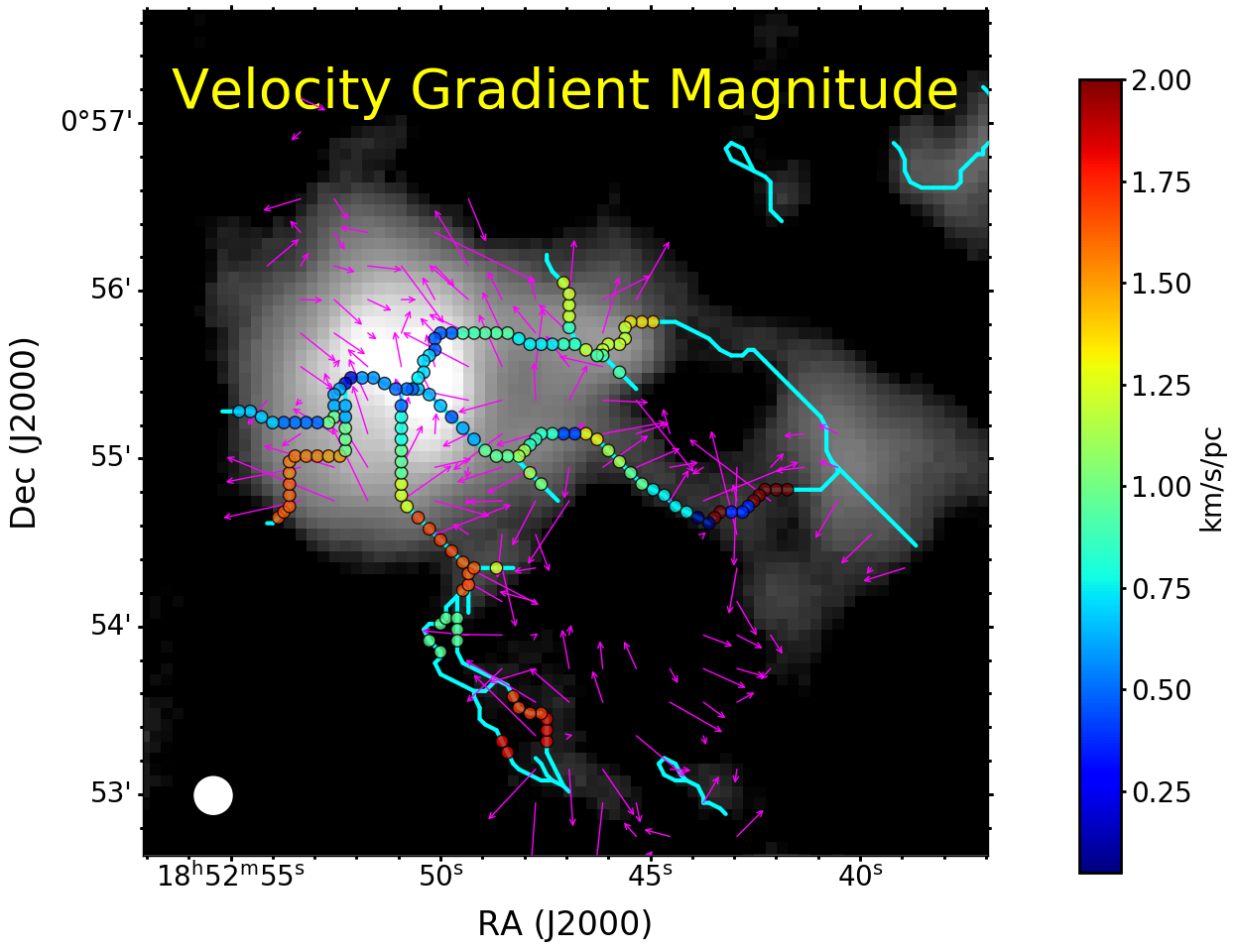}
\caption{Magnitude of local LOS velocity gradient along filaments overlaid on the JCMT 850 $\mu$m continuum map. The velocity gradient magnitude is higher ($\sim$1--2 km/s/pc) in the outer area and  decreases to $\sim$0.5 km/s/pc near the hub center. }\label{fig:vga_map}
\end{figure}

\begin{figure}
\includegraphics[width=\columnwidth]{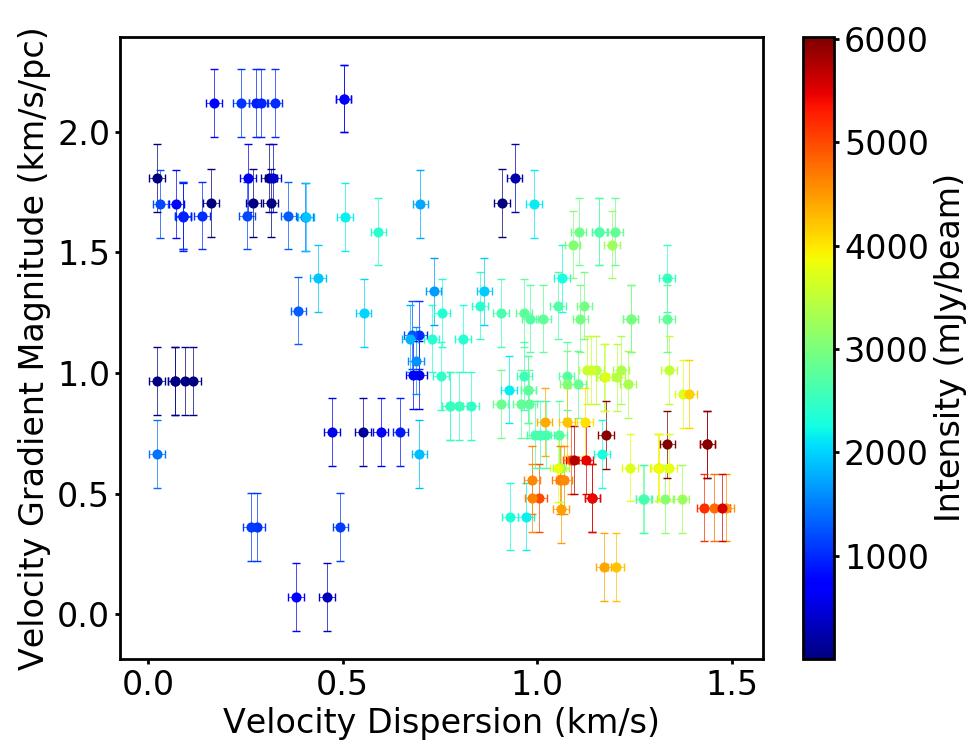}
\caption{Magnitude of local LOS velocity gradient versus LOS velocity dispersion. Color-coded is the 850 $\mu$m intensity. A higher intensity indicates regions closer to the hub center. As the intensity increases, the velocity gradient magnitude decreases while the velocity dispersion increases. }\label{fig:vg_vd}
\end{figure}

In addition to the direction of the LOS velocity gradient, we find that the magnitude of the velocity gradient decreases from the outer area to the central hub (\autoref{fig:vga_map}). Since the differential orientations between filaments and local velocity gradient tend to be overall random, the local velocity gradient components (parallel and perpendicular to filaments) both also decrease in magnitude towards the central hub. In addition, we further find that the LOS velocity dispersion increases while the local velocity gradient magnitude decreases, as illustrated in \autoref{fig:vg_vd}.

The opposite trend between local velocity gradient magnitude and velocity dispersion may suggest a change in the velocity power spectrum. It is known that turbulence and the correlated gas kinematics are scale-dependent in molecular clouds, as described by Larson's law \citep[e.g.,][]{la81,mc07}. Observationally, LOS velocity gradients are often calculated from LOS velocities measured at pixels with particular separations, usually the beam size. Adopting a centroid velocity only traces the velocity of the densest component along the line of sight, but lacks the information of velocity variation along the line of sight. Hence, a resulting velocity gradient is only sensitive to the gas kinematics at a scale comparable to the physical distance between these densest components of the selected pixels. Since G33.92+0.11 is likely a face-on system \citep{liu12}, the physical distance is approximately the plane-of-sky distance. Hence, the estimated local velocity gradients are sensitive to gas motions on a $\sim0.5$ pc scale. In contrast, the observed velocity dispersion includes velocity information of all components at all physical scales along the line of sight. 

An increase in velocity dispersion towards high-density regions has been observed in the HFS SDC13 \citep{wi18}. Since turbulent energy is expected to dissipate in stagnation regions generated by shock compression, contrary to the observed trends (\autoref{fig:vg_vd}), the growing velocity dispersion more likely originates from gravity instead of turbulence. \citet{li19} have simulated the formation of filaments in a magnetized cloud driven by turbulence and self-gravity. Their velocity power spectra reveal an increase in power at small scales ($\lesssim0.1$ pc) after gravity is turned on.  Self-gravity also causes growing power of density structures at a $\lesssim0.5$ pc scale. If observed in low resolution ($\gtrsim0.5$ pc), this subparsec-scale 
processes
could result in an increase in velocity dispersion, as seen in SDC13 and also here in G33.92+0.11.

The high-resolution SMA and ALMA molecular line observations reveal that the inner 0.5 pc area of the G33.92+0.11 central hub is fragmented into several young clusters surrounded by four converging spiral-arm-like structures \citep{liu12,liu15}. About 30 young stellar objects are identified within these clusters \citep{liu19}. These structures suggest that local gravitational fragmentation has taken place within the central hub of G33.92+0.11. These 0.01-pc scale fragmentation process is similar to the \citet{li19} model, and hence might explain the increase in velocity dispersion seen with our resolution (0.5 pc). In addition, the observed spiral-arm-like morphology and the related virial analysis suggest that the rotational energy is important in supporting these 0.1 pc scale structures \citep{liu19}. The accelerated rotational motion near the hub center, due to the conservation of angular momentum, could then also contribute to the increase in velocity dispersion. Furthermore, as the central hub of G33.92+0.11 hosts an ultracompact H\Rom{2} region, the feedback from the massive star formation may further enhance the turbulence within the hub, which has been seen in \citet{liu12,liu19}. 


A decreasing velocity gradient magnitude towards high-density regions is rarely seen in other systems. The opposite trend has been seen in
a few systems, such as S242 \citep{yu20}, Mon R2 \citep{tr19}, and two of the filaments in SDC13 \citep{wi18}. On the other hand, a constant velocity gradient magnitude, independent of local density, has also been observed, such as in G35.39 \citep{so19} and the other two filaments in SDC13 \citep{wi18}. An increased velocity gradient magnitude toward  high-density regions is expected for accretion flows due to gravitational acceleration \citep[e.g.,][]{tr19}, while an opposite trend can likely be related to the gas dynamics originating from large-scale turbulence (e.g., the turbulent fragmentation). This is because large-scale turbulence is expected to dissipate in the high-density stagnation regions. It is still unclear which trend is more common, as the variation of local velocity gradient magnitudes has only been studied in a few systems. To make matters more complicated, the current observations have been probing the velocity gradient at different physical scales, due to the different tracers, angular resolutions, and object distances. A larger and more homogeneous sample with multi-scale velocity gradient measurements is still needed to understand the origin of the magnitude variations.

\subsection{Origin of the HFS}\label{sec:originHFS}
Our virial analysis, accounting for the global energy balance of G33.92+0.11, shows that the gravitational energy overall dominates the kinematic and magnetic energy in this system, and hence the whole system is expected to contract. Locally, we find that the filament orientations tend to be aligned with local gravity. This suggests that these filaments are likely accretion streams driven by gravity. In addition to the global contraction, the local velocity gradients reveal gas motions perpendicular to the major filaments in the diffuse areas, hinting that these filaments
possibly accumulate ambient gas from the diffuse surrounding while also flowing toward the hub center. 

\begin{figure*}
\includegraphics[width=\textwidth]{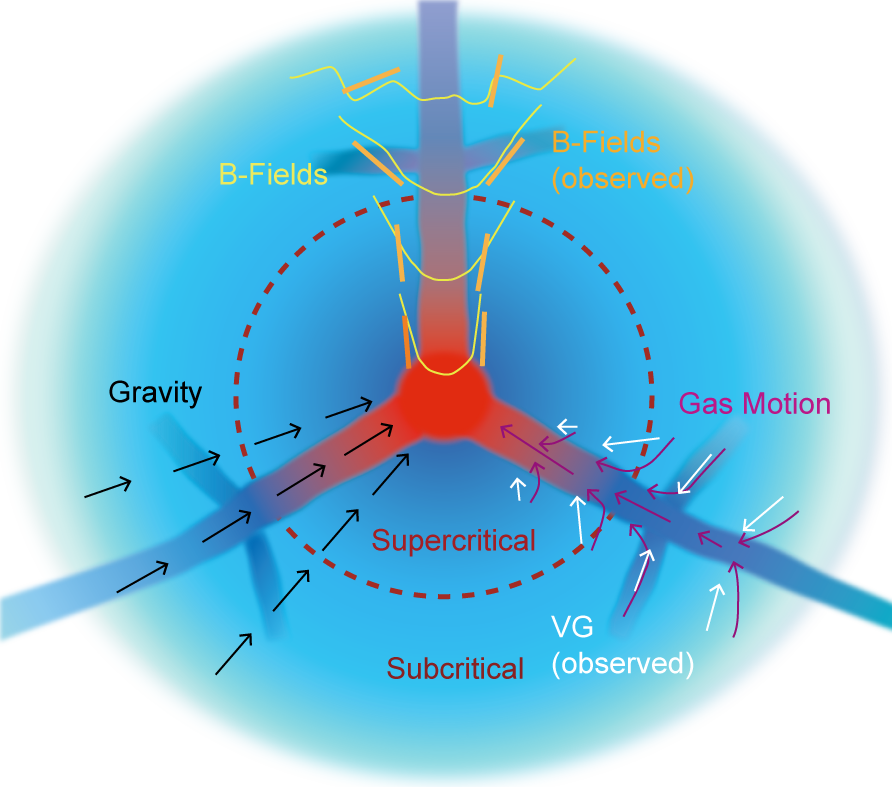}
\caption{Cartoon illustrating observed features.
The black arrows represent the directions of local gravity. The yellow curve shows a model-compatible
magnetic field morphology (section \ref{sec:originHFS}) with the orange segments displaying the observed field segments with a spatial resolution ($\sim$0.5 pc) comparable to the filament widths (0.5--1 pc). The magenta arrows illustrate the directions of 
gas motion with the white arrows depicting the observed velocity gradients at the resolved 0.5-pc scale. The background color displays local density. In the diffuse outer areas, the ambient gas is accumulated onto the major filaments directly or via short streams, while the major filaments flow toward the central hub due to global gravity. Magnetic fields might be initially perpendicular to the major filaments, guiding the gas accumulation, but are then also dragged by the gas flows along the main filament. 
In the dense inner areas, the density of the major filaments grows through gathering mass from the surroundings. The increased ram pressure of the accretion flows can further stretch the magnetic field on the way to the center. This results in a magnetic field parallel to both the major filaments and local gravity. 
The magnitude of the velocity gradient decreases towards the center while the velocity dispersion increases. 
An outer subcritical and inner supercritical zone is indicated. 
}\label{fig:cartoon}
\end{figure*}

The role of the magnetic field is relatively less important in this system, as compared to gravity and turbulence. In the dense regions, both filaments and magnetic fields are aligned with the gravitational force. Hence the magnetic tension force cannot retard the accretion streams effectively, although it can still help to stabilize the filaments against radial collapse by increasing the critical linear density. 
In the diffuse regions, magnetic fields still tend to be aligned with the major filaments, but perpendicular to those minor filaments merging into the major filaments. This suggests that magnetic fields might be important in delaying the mass accumulation onto the major filaments. Hence, the density growth rate of the major filaments is less than their turbulence growth rate. Consequently, the turbulent pressure might keep supporting the major filaments from radial collapse until approaching the center. This can then result in a massive protocluster forming in the hub center, instead of numerous young stars randomly distributed over the system. Our interpretation based on all the observed features is schematically illustrated in \autoref{fig:cartoon}.

This interpretation provides a possible explanation to the question of how massive stars/clusters form in a globally subvirial ($5\sigma_v^2R/(GM)<2$) cloud, where the turbulence pressure is insufficient to support the cloud against gravity, and the cloud is, therefore,  expected to fragment to form numerous lower-mass stars before accumulating sufficient mass for massive star formation \citep{mc03}. \citet{ka13} argue that the additional support from magnetic fields might stabilize such subvirial clouds and enable massive star formation. However, our virial analysis shows that both magnetic and kinematic energy are insufficient to globally support G33.92+0.11 against gravity, but yet massive star formation is still present. Our analysis further shows that a \textit{globally} unstable system might still host \textit{locally} stable filaments, through which mass can converge toward the center -- without fragmenting into low-mass clumps -- and enable massive star formation.
These findings illustrate the importance of jointly
studying global and local properties in order
to understand such systems.

Our findings favor the multi-scale gravitational collapse cloud model in \citet{go14}. In this model, a super-Jeans cloud forms from colliding flows and rapidly begins to undergo gravitational collapse. The collapse soon becomes nearly pressureless, proceeding along its shortest dimension, and forms filamentary sub-structures. The resulting filaments are not in a static equilibrium but are long-lived flow structures that accumulate ambient gas from their environment and direct it towards the major gravitational potential well (hub center). This model is based on the physical condition that both thermal and turbulent pressure of the forming cloud are insufficient to stabilize the cloud against gravitational collapse. This is consistent with our estimated gravitational energy that dominates the kinematic energy. The observed increasing filament ridge intensities from outer to inner areas can be explained by the accretion flows accumulating nearby gas while at the same time continuously moving toward the center. However, we do not detect increasing velocity gradients along the filaments as they approach the center. This is possibly because we do not have sufficient resolution to resolve the velocity structures within the filaments. \citet{liu12} have detected a velocity gradient of 0.96 km/s/pc along the filaments at $\sim$ 0.1 pc scale within the inner 0.3 pc area of this system using SMA NH${_3}$ (1,1) and (3,3) line data, which is higher than the velocity gradient of $\sim$0.5 km/s/pc that we find in the hub center. This might support our speculation that the gas within the filaments is gravitationally accelerated near the center.

In the above emerging scenario, magnetic fields are expected to be dragged by the collapsing gas as simulated in \citet{go18}, a model identical to \citet{go14} but with magnetic fields included. As a result, the magnetic fields are perpendicular to filaments in diffuse areas because the fields are dragged by the gas accreted onto the filaments. As density increases, the magnetic field lines become parallel to filaments because the fields are stretched by the longitudinal flow along the filaments. Our observed alignment between magnetic fields and filaments in dense areas supports this prediction. 
Our finding for the diffuse areas is clearly revealing a distribution different
from the one for dense areas, being more random
but also hinting a broader peak towards 
larger misalignments, i.e., field orientations perpendicular to filaments. 


\section{Conclusions}\label{sec:con}
This paper presents the 850 $\mu$m JCMT SCUBA-2/POL-2 observations toward the G33.92+0.11 hub-filament system. Our observations reveal 
an organized but complex 
magnetic field morphology. 
From the analysis of these data, we find the following results.

\begin{itemize}
    \item Filamentary structures are identified in G33.92+0.11 using the $DisPerSE$ algorithm. The identified filaments appear as an overall converging network where most of the filaments either directly connect to the central hub or merge with other filaments connecting to the central hub. The filament converging points are generally located at local intensity peaks.
    
    \item The direction of local gravity, estimated from the 850 $\mu$m continuum data, mostly points toward the central hub and is prevailingly aligned with the filament orientations. This provides some evidence that the filaments in this system are controlled by gravity. These filaments have widths varying from $\sim$0.5 to $\sim0.8$ pc as a function of the ridge intensity. This is different from the constant filament width seen in $Herschel$ surveys.
    
    \item We further analyze the relative orientations between filaments, local gravity, magnetic field, and local LOS velocity gradients estimated from IRAM-30m \C18O (2-1) line data. The randomness of the distributions of these relative orientations is examined using KS-tests with a 95\% confidence interval. The resulting statistics indicate that local gravity is aligned with filaments and the magnetic field in high-density regions. In low-density regions, the filaments are still aligned with local gravity, but
    shifting to becoming more perpendicular or
    random to the magnetic field and velocity gradients.
    
    \item Globally, the relative importance of gravitational, magnetic, and kinematic energy in G33.92+0.11 is estimated from a Virial equation. The analysis suggests that the gravitational energy dominates the system, while the kinematic is slightly larger than the magnetic energy by a factor of $\sim$1--2. Hence, the system is likely collapsing, and the identified filaments are likely infalling accretion flows.
    
    \item Combining all findings from global properties and local correlations, we interpret G33.92+0.11 as a multi-scale gravitationally collapsing cloud with relatively weak turbulence and magnetic field. The ambient gas in the diffuse environment is accreted onto the filaments, while the filaments drag the magnetic field lines and flow toward the gravitational center. The observed local velocity gradients mainly trace the gas accumulation from the surrounding to the filaments, especially in low-density areas, and are thus perpendicular to the filaments. The observed magnetic field is stretched by the accretion flows, especially in high-density areas, and is therefore aligned with filaments and gravity.
    
\end{itemize}

\acknowledgments
The James Clerk Maxwell Telescope is operated by the East Asian Observatory on behalf of The National Astronomical Observatory of Japan; Academia Sinica Institute of Astronomy and Astrophysics in Taiwan; the Korea Astronomy and Space Science Institute; the Operation, Maintenance and Upgrading Fund for Astronomical Telescopes and Facility Instruments, budgeted from the Ministry of Finance (MOF) of China and administrated by the Chinese Academy of Sciences (CAS), as well as the National Key R\&D Program of China (No. 2017YFA0402700). Additional funding support is provided by the Science and Technology Facilities Council of the United Kingdom and participating universities in the United Kingdom and Canada. Additional funds for the construction of SCUBA-2 and POL-2 were provided by the Canada Foundation for Innovation. The Starlink software \citep{cu14} is currently supported by the East Asian Observatory. The authors wish to recognize and acknowledge the very significant cultural role and reverence that the summit of Maunakea has always had within the indigenous Hawaiian community. We are most fortunate to have the opportunity to conduct observations from this mountain. J.W.W and P. M. K. acknowledge support from the Ministry of Science and Technology (MOST) grants 
MOST 108-2112-M-001-012
and MOST 109-2112-M-001-022,
and from an Academia Sinica Career Development Award. S.P.L. is thankful to the support from the Ministry of Science and Technology of Taiwan through the grant MOST 106-2119-M-007-021-MY3. R.G.M. acknowledges support from UNAM-PAPIIT project IN104319.

\appendix
\section{Polarization Properties}\label{sec:app_pp}
\begin{figure*}
\includegraphics[width=\textwidth]{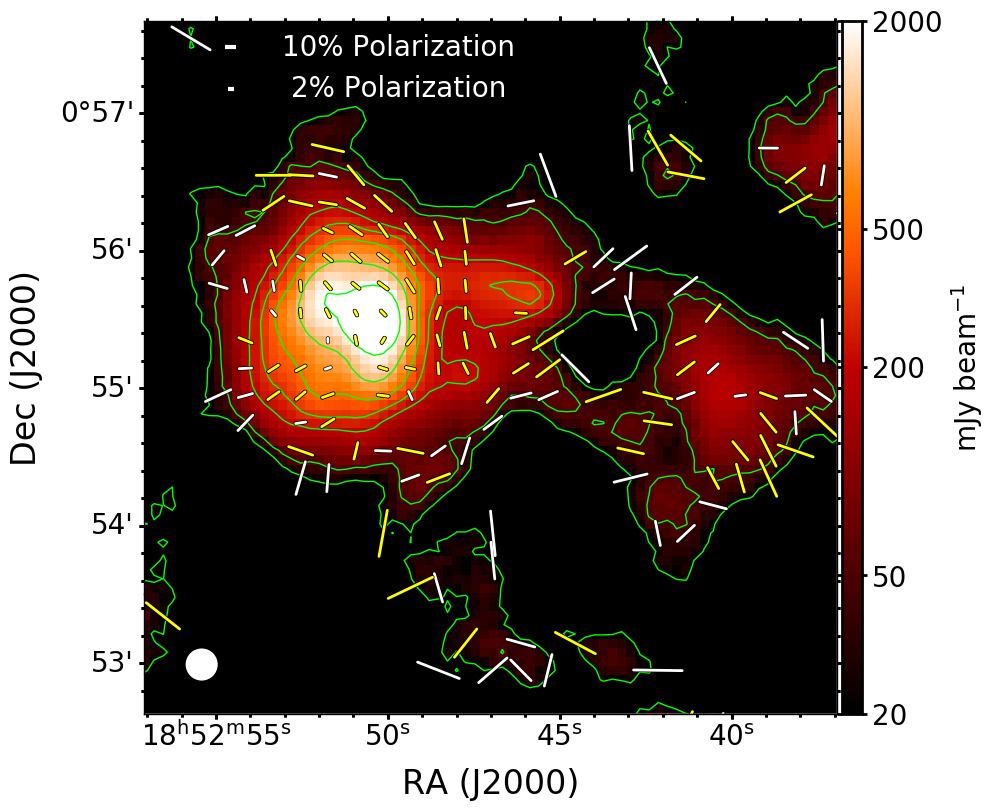}
\caption{Polarization (segments) sampled on a 12\arcsec\ grid overlaid on 850 $\mu$m dust continuum
(color and contours), sampled on a 4\arcsec\ grid, of the G33.92+0.11 region. The segments are selected with the criteria $I/\sigma_{I}>10$ and $P/\sigma_{P}>2$. The yellow and white segments display the larger than 3$\sigma$ and 2--3$\sigma$ polarization detections. The lengths of the segments are proportional to the square root of the polarization fraction. The white segments in the upper left corner are examples of 2\% and 10\% polarization fractions. The green contours show the total intensity at 20, 50, 200, 300, 500, 1000, and 2000 m\jyb. The white circle in the bottom left corner is the JCMT beam size of 14\arcsec. The rms noise of the Stokes I is $\sim$1--5 m\jyb\, depends on pixel intensities, within the central 3\arcmin\ area. 
}\label{fig:pmap_raw}
\end{figure*}

\autoref{fig:pmap_raw} shows the observed polarization map overlaid on the 850 $\mu$m total intensity images.
Assuming that dust grains are aligned with the magnetic field is 
a fundamental assumption that allows us to use polarization data to trace magnetic fields. 
The relation between 
the total intensity (I) and the polarization fraction
(P) is commonly used to examine this assumption. If the dust grains are not aligned with the magnetic field in a dense cloud, the observed polarized intensity (PI) would be independent of the cloud's column density, and thus $PI\propto I^{0}$ or $P=PI/I\propto I^{-1}$. In contrast, if magnetic-field-aligned dust grains are present in a dense cloud, we expect an increase of PI with the cloud's column density, and thus $PI\propto I^{\alpha}$ or $P\propto I^{-\alpha}$, where $\alpha$ is smaller than 1.

\begin{figure}
\includegraphics[width=\columnwidth]{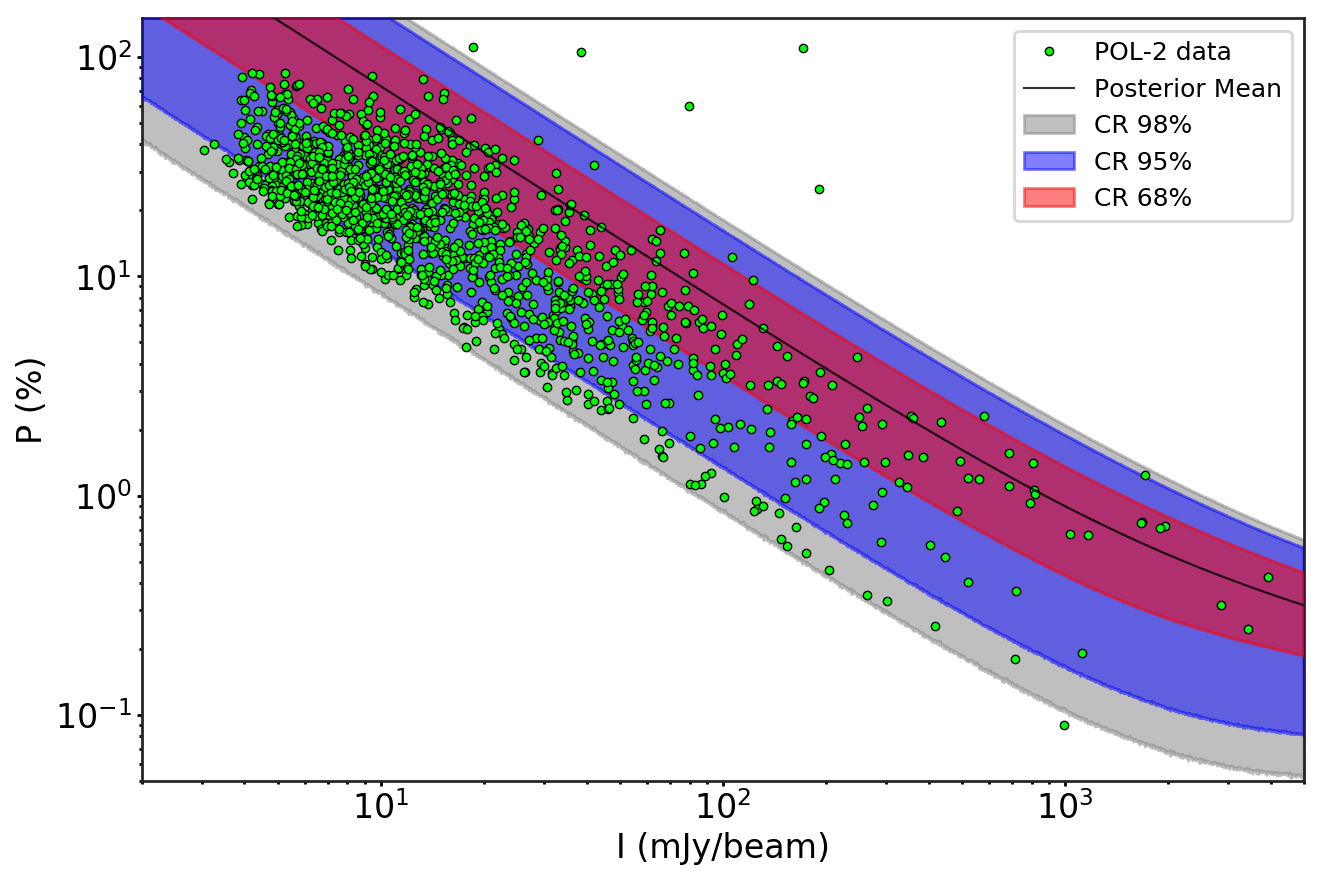}
\caption{850 $\mu$m total intensity $I$ vs polarization fraction $P$. The green points are the non-debiased, $I/\sigma_{I}>$3, and $\sigma_I<$5 m\jyb\ POL-2 polarization measurements. The colored regions are the predicted I-P distributions based on the Bayesian analysis within the 68\%, 95\%, and 98\% confidence region (CR). The black line indicates the posterior mean. Most of the data points are within the 98\% confidence region of our prediction.}\label{fig:IP}
\end{figure}

The observed I--P relation for G33.92$+$0.11 is displayed in \autoref{fig:IP}. We follow the Bayesian analysis in \citet{wa19} to determine the power-law index $\alpha$ of this relation: The model we use is a power-law
\begin{equation}\label{eq:prior}
P =\beta I^{-\alpha},
\end{equation}
with a probability distribution function (PDF) of P described by the Rice distribution
\begin{equation}\label{eq:rice}
F(P|P_0)=\frac{P}{\sigma_P^{2}}exp\left[-\frac{P^2+P_0^2}{2{\sigma_P}^2}\right]I_0\left(\frac{PP_0}{\sigma_P^2}\right),
\end{equation}
where $P$ is the observed polarization fraction, $P_0$ is the real polarization fraction, $\sigma_P$ is the uncertainty in the polarization fraction, and $I_0$ is the zeroth-order modified Bessel function. We further assume that the uncertainty in the polarization fraction is given by
\begin{equation}
\sigma_P=\sigma/I,
\end{equation}
where $\sigma$ is the uncertainty of Stokes Q and U.
We, nevertheless, 
treat $\sigma$ as an unknown parameter here
because the dispersion in the observed $P$ is not only coming from the observational uncertainty but also from the intrinsic dispersion within G33.92+0.11.

\begin{figure}
\includegraphics[width=\columnwidth]{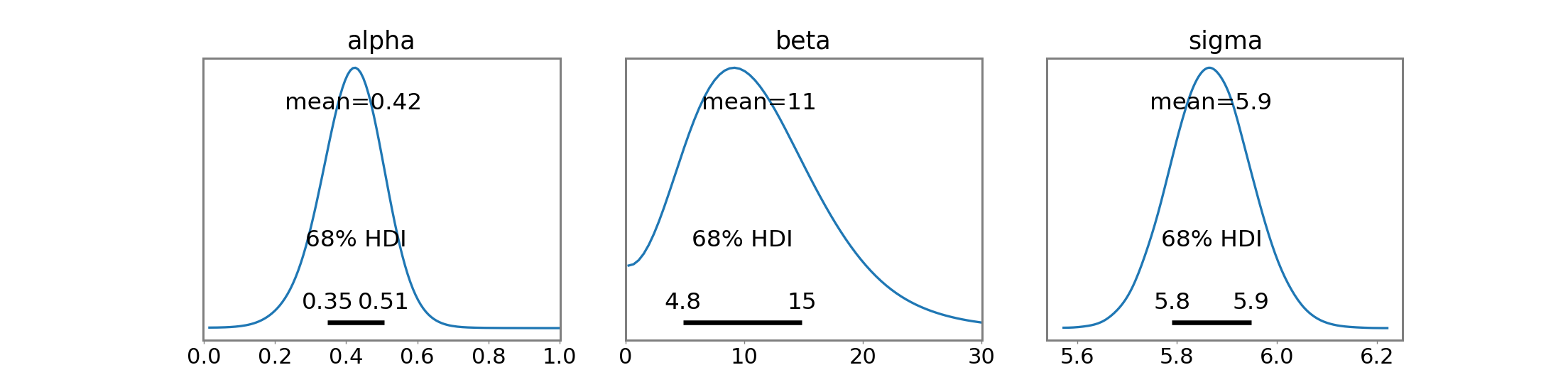}
\caption{Posterior distributions from Bayesian analysis for I--P relation. The black lines delineate the 68\% ($1\sigma$) highest density interval (HDI). $\alpha$ is expected to be significantly less than 1 with $0.42\substack{+0.09 \\ -0.07}$, suggesting that aligned dust grains are present within G33.92+0.11. }\label{fig:IPpost}
\end{figure}

Since the distribution of the Ricean noise is well accounted for in this model, we can directly use the non-selected, non-debiased polarization data to compare with the model. Hence, in \autoref{fig:IP}, we use all the raw polarization data for the Bayesian analysis, except for the edge pixels or the possible artificial patterns selected by the criteria $\sigma_I > 5$ m\jyb\ and $I/\sigma_I>3$. The computed posterior distributions of the Bayesian analysis are shown in \autoref{fig:IPpost}. $\alpha$ is expected to be $0.42\substack{+0.09 \\ -0.07}$ which is significantly smaller than 1. This suggests that aligned dust grains are present within G33.92+0.11, and hence the observed polarization patterns likely trace the magnetic field morphology.

\section{Pairwise Differential Orientations between Filaments, Magnetic field, Local Gravity, and Local Velocity Gradients}\label{sec:full_cor}
\begin{figure*}
\includegraphics[width=\textwidth]{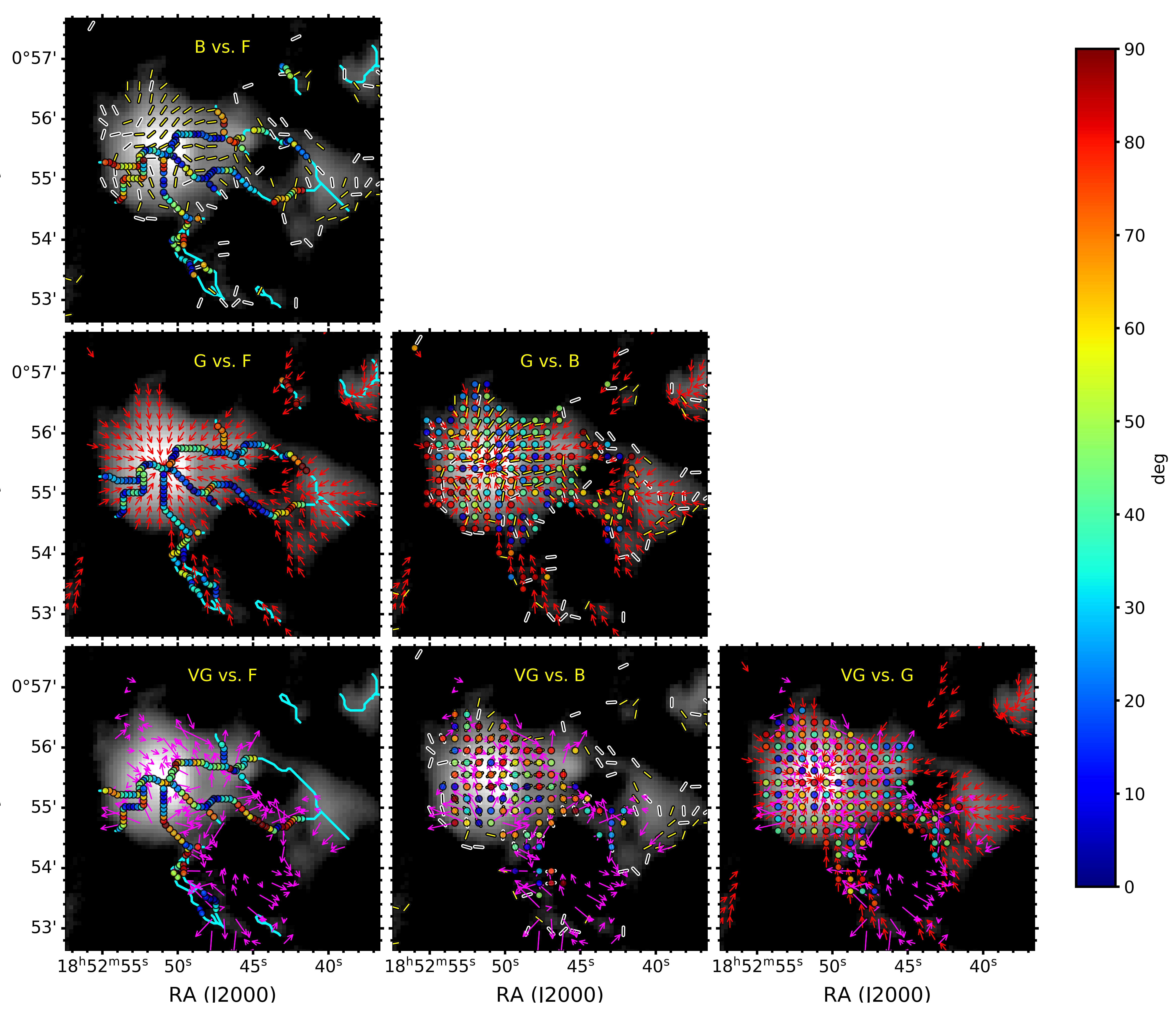}
\caption{Pairwise differential orientation maps between filaments, magnetic field, local gravity, and local velocity gradients, overlaid on 850 $\mu$m dust continuum. The cyan lines are the filaments identified in \autoref{fig:fi}. The yellow and black segments represent the magnetic field orientations as shown in \autoref{fig:pmap}, the red arrows are the projected local gravity, and the magenta arrows show the local velocity gradients. 
Filled color-coded circles (color wedge) are the pairwise differential orientations $\Delta PA$.
}\label{fig:dpa_map_all}
\end{figure*}

\autoref{fig:dpa_map_all} presents all combinations of pairwise differential orientations between filaments, magnetic field, local gravity, and local velocity gradients, spatially overlaid on the 850$\mu$m dust continuum map. The first column is identical to \autoref{fig:dpa_map}. The VG--B and VG--G plots might display tendencies of the local velocity gradient being perpendicular to local gravity and magnetic field away from the hub center, except for some regions located in between filaments. However, our spatial resolution is insufficient to probe this possible change from
on-filament to off-filament regions in more detail. The G--B plot generally shows that the magnetic field is aligned with local gravity near the hub center. Only a few exceptions are found at the northern side of the C1 converging point. 
These exceptions are at the center of two filaments, where we find that the magnetic field orientations change significantly (from PA$\sim$ 135\degr\ to 90\degr\ at C1, and from PA$\sim$10\degr\ to 90\degr\ at C3). Hence, we speculate that the observed magnetic field structure in theses areas is possibly the result of the merging of two 
nearby filaments, and thus the alignment between magnetic field and local gravity is perturbed.

\section{Dependence of Filaments, Magnetic field, Local Gravity, and Local Velocity Gradients on Intensity}\label{sec:dpa_I}
\begin{figure*}
\includegraphics[width=\textwidth]{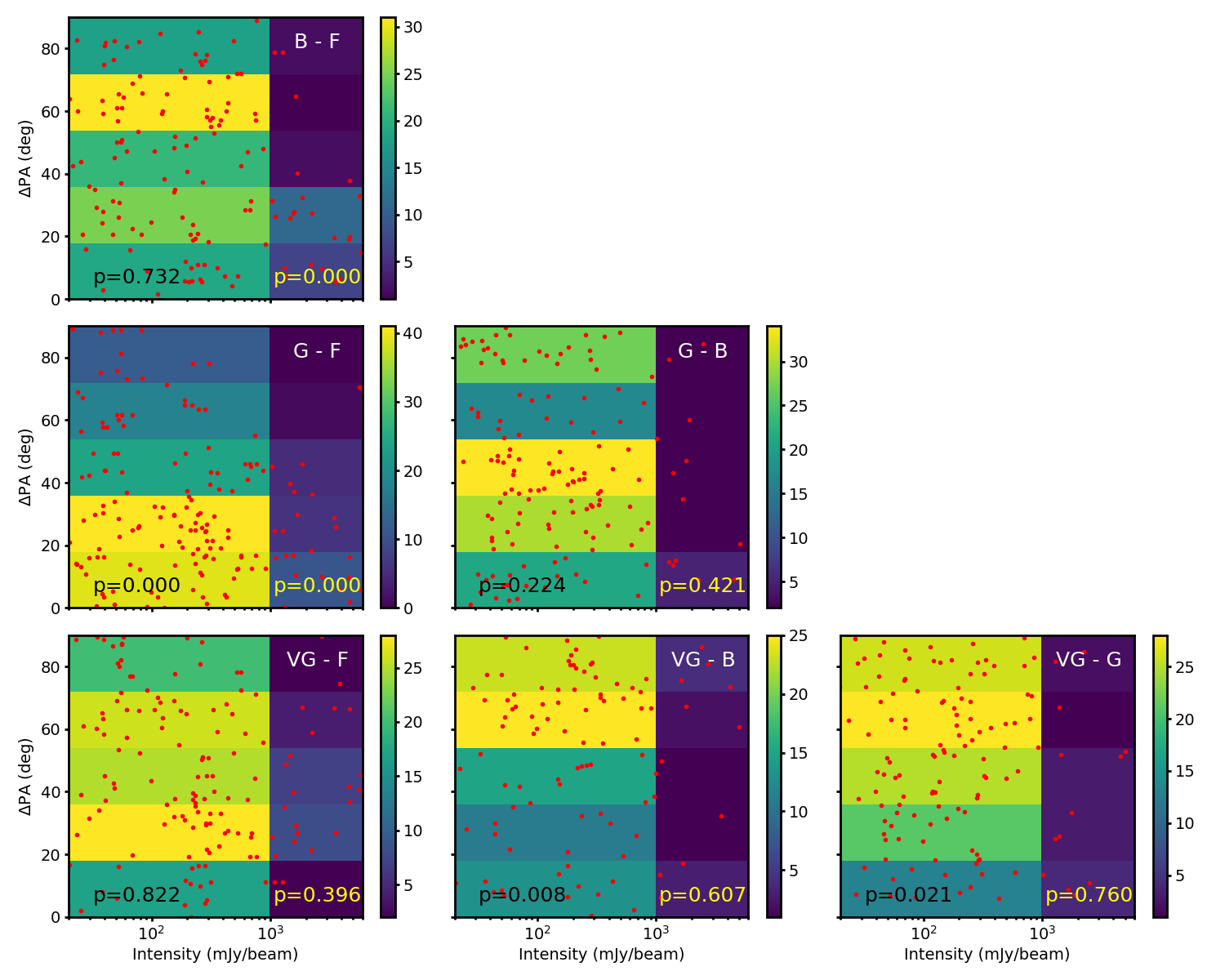}
\caption{Pairwise differential orientations between filaments, magnetic field, local gravity, and local velocity gradients versus intensity. The red points represent the $\Delta PA$ of the pairwise parameters (indicated in the upper right corner in each panel)
as a function of
the local intensity at spatially matched pixels/segments/vectors. 
The background color indicates the density of 
red data points given with the color wedge.
The data set in each panel is separated into a
low- and high-intensity group, below and above $I_{850}=1000$ m\jyb. KS tests are applied 
and p-values are given for 
each group separately.
The comparison between the $\Delta PA$ distributions in the low- and high-intensity areas reveals a tightening alignment between filaments, gravity, and magnetic fields with growing intensity, while the local velocity structures become more disordered.}\label{fig:dpa_I}
\end{figure*}

\autoref{fig:dpa_I} shows the distribution of $\Delta$PA and total intensity for all the associated filaments, magnetic field, local gravity, and local velocity gradient pairs. The data set in each panel is separated into a low- and high-intensity group, below and above $I_{850}=1000$ m\jyb. The B--F, VG--B, and VG--G pairs show different $\Delta PA$ distributions between low- and high-density areas, which are also shown in \autoref{fig:hist_dpa_2I}. We note that the high-density groups mostly contain only 10--30 data points, and thus the high p-value can be due to the small sample sizes.

\section{Filament Identification Threshold }\label{sec:f20}
\begin{figure}
\includegraphics[width=\columnwidth]{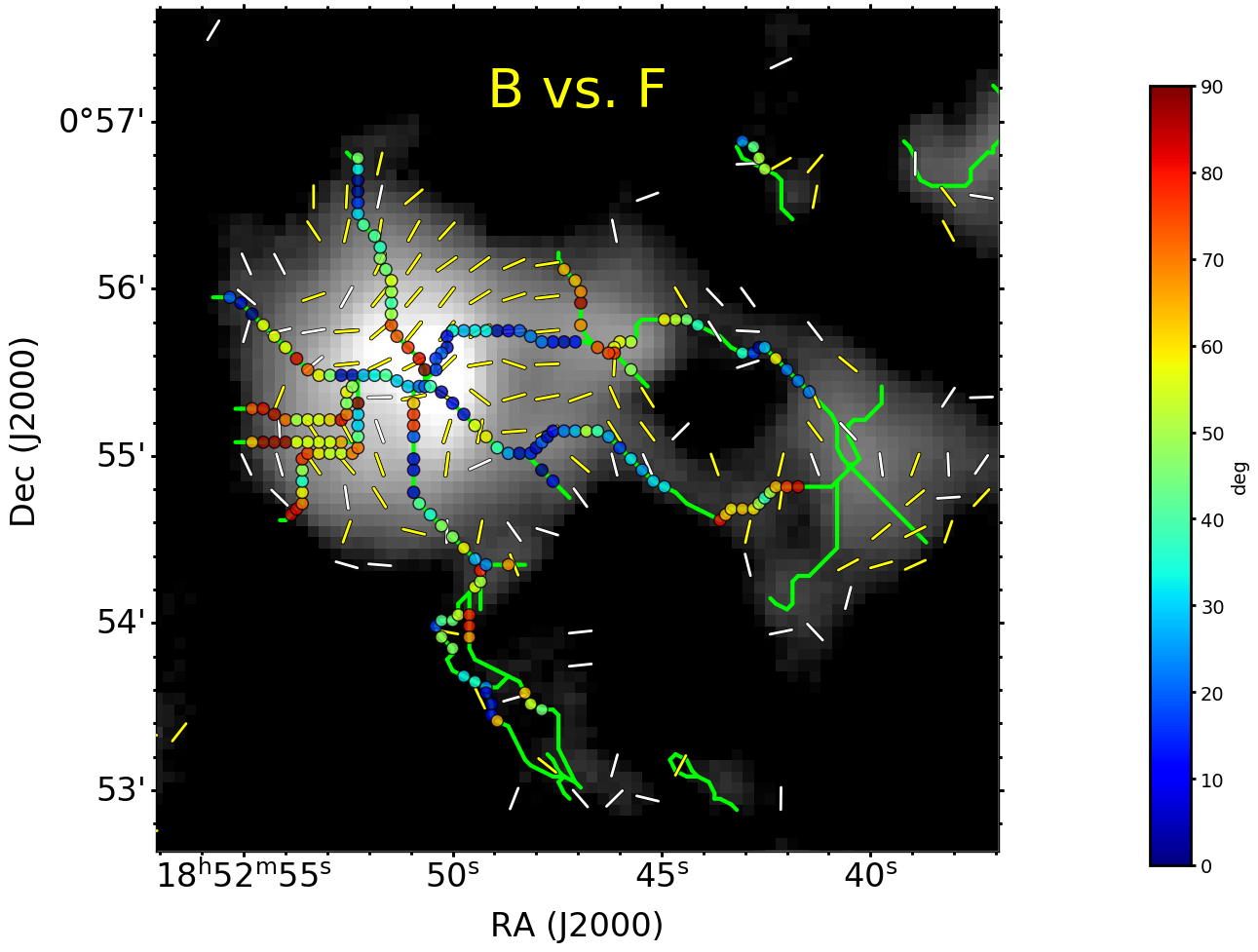}
\caption{Differential orientation map for filaments (F) vs magnetic fields (B), where the filaments are identified with a lower threshold of 20 m\jyb. The yellow and white segments are the magnetic field orientations as shown in \autoref{fig:pmap}. 
The color-coded circles are the differential orientations B vs F. 
}\label{fig:dpa_map_f20}
\end{figure}

\begin{figure}
\includegraphics[width=\columnwidth]{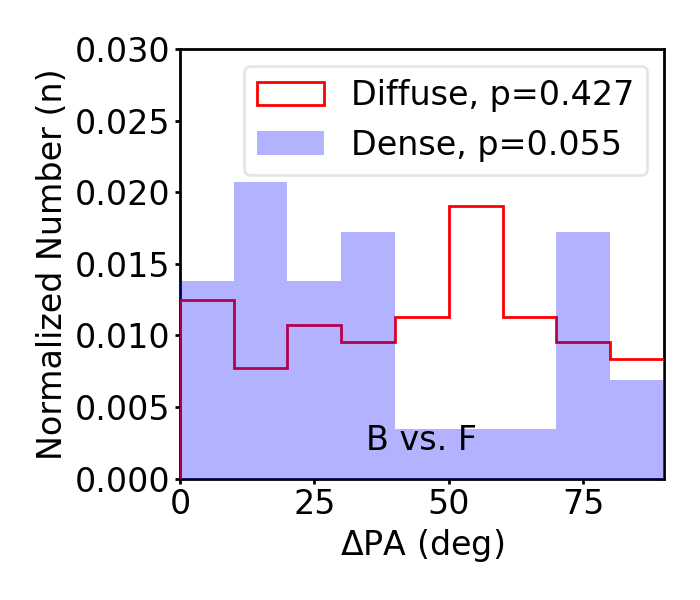}
\caption{Normalized histograms of differential orientations between filaments, identified with a lower threshold of 20 m\jyb, and magnetic field, separated by local intensity. The blue and red histograms represent the data selected in the dense ($I_{850}>$ 1000 m\jyb) and the diffuse ($I_{850}<$ 1000 m\jyb) areas. The KS-test results against a random distribution are listed for these two groups.}\label{fig:hist_dpa_f20}
\end{figure}

In order to investigate how a filament identification threshold possibly affects our results in \autoref{sec:4par} (\autoref{fig:dpa_map}, \autoref{fig:hist_dpa}, and \autoref{fig:dpa_I}), we have repeated our analysis
with a lower threshold of 20 m\jyb.
The KS-tests still reveal correlations identical
to the results in \autoref{sec:4par} for a higher threshold. 
The only exception is the p-value for the B--F relation in dense regions. This value increases from $<0.001$ to 0.055, as a result of a histogram originally peaking around small values (top panel 
in \autoref{fig:hist_dpa_2I}) now shifting to a more bimodal $\Delta PA$ distribution around 0\degr\ and 90\degr\ (\autoref{fig:hist_dpa_f20}).
These more perpendicular B--F pairs can be traced back to a a newly located weaker filament extending from the central hub to the north (\autoref{fig:dpa_map_f20}). 
Around this location, the magnetic field seems to be more aligned with the nearby filament extending toward the west, and not with the new filament toward the north. 
Since these two filaments merge into the central hub and are separated by only $\sim$0.5--1 pc (less than two times the filament width of 1--1.5 pc), it is likely that the observed polarization here mainly traces the magnetic field around the denser filament to the west. These leads to the few perpendicular B--F pairs. However, it has to be acknowledged that the limited resolution does not really allow us to
fully separate these two filaments near the center. We, therefore, conclude that our results from \autoref{sec:4par} remain robust and largely unaffected if the detection threshold for filaments is lowered to 20 m\jyb.


\vspace{5mm}
\facilities{JCMT,IRAM:30m}

\software{Aplpy \citep{aplpy2012,aplpy2019}, Astropy \citep{astropy2013,astropy2018}, DisPerSE \citep{so11}, FilChap \citep{su19}, GILDAS/CLASS \citep{pe05,gi13}, NumPy \citep{numpy}, SciPy \citep{scipy}, PySpecKit \citep{gi11}, Smurf \citep{be05,ch13}, Starlink \citep{cu14}}
\newpage
\bibliography{references}{}
\bibliographystyle{aasjournal}
\end{document}